%% file: main.tex
\documentclass[twocolumn]{article}
\usepackage{parskip}
\usepackage{amssymb}
\usepackage{amsmath}
\usepackage{fullpage}
\usepackage{authblk}
\usepackage{blindtext}
\usepackage{siunitx}
\usepackage{float}
\usepackage{caption}
\usepackage[superscript,nomove]{cite}
\usepackage{hyperref}
\usepackage[normalem]{ulem}
\usepackage{graphicx}
\usepackage{cancel}
\usepackage{pifont}
\usepackage{diagbox}
\usepackage[percent]{overpic}
\usepackage{tikz}
\usepackage{stfloats}
\usepackage{cuted}
\usetikzlibrary{calc}
\usetikzlibrary{positioning}
\usetikzlibrary{arrows.meta}

\makeatletter \renewcommand{\@citess}[1]{\textsuperscript{[#1]}} \makeatother

\makeatletter
\def\@firstoftwo@second#1#2{%
  \def\temp##1.##2\@nil{##2}%
   \temp#1\@nil}
\newcommand\sref[1]{%
   (A.\expandafter\@setref\csname r@#1\endcsname\@firstoftwo@second{#1})%
}

\makeatother

\title{First-principles transition-state tensorial cluster expansion of vacancy diffusion in Ta-W beyond the kinetically-resolved activation approximation}
\author[1]{\small Jacob Jeffries \thanks{jwjeffr@g.clemson.edu}}
\author[2]{Brianna Sebastian-Olazabal}
\author[1,2]{Enrique Martinez \thanks{enrique@clemson.edu}}
\affil[1]{Department of Materials Science and Engineering, Clemson University, Clemson, SC 29634, USA}
\affil[2]{School of Mechanical and Automotive Engineering, Clemson University, Clemson, SC 29634, USA}
\date{\small \today}

\renewenvironment{abstract}
 {\quotation\small\noindent\rule{\linewidth}{.5pt}\par\smallskip
  {\centering\bfseries\abstractname\par}\medskip}
 {\par\noindent\rule{\linewidth}{.5pt}\endquotation}

\begin{document}

\twocolumn[
  \begin{@twocolumnfalse}
  \maketitle
    \begin{abstract}
        Predicting diffusion in chemically complex alloys remains challenging due to the strong dependence of migration barriers on local atomic environments. Migration barriers computed using density functional theory and nudged elastic band calculations are represented via a tensorial cluster expansion including transition states and deployed in on-lattice kinetic Monte Carlo simulations. Applied to the Ta-W system, the framework captures nontrivial composition-dependent diffusion behavior arising from a crossover between solute trapping and percolated low-barrier transport pathways, yielding a maximum in the apparent activation energy near intermediate compositions. This approach establishes a general and scalable route for integrating first-principles transition-state energetics into mesoscale kinetic simulations, enabling predictive multiscale modeling of diffusion in chemically complex materials and providing a pathway for uncovering emergent transport phenomena.
    \end{abstract}
  \vspace{0.5cm}
  \end{@twocolumnfalse}
]

\input{sections/introduction}
\input{sections/model}
\input{sections/methods}
\input{sections/discussion}
\input{sections/limitations}
\input{sections/conclusions}
\input{sections/data_availability}
\input{sections/acknowledgements}
\input{sections/disclaimer}
\input{sections/appendix.tex}

\bibliography{bibfile.bib}

\end{document}

%% file: sections/introduction.tex
\section{Introduction}

High-throughput screening of concentrated alloys is necessary for the discovery and development of damage-resistant materials for clean energy applications. Primarily, the effects of radiation on materials are driven by point defect (PD)-mediated diffusion, underpinning phase stability, precipitation, segregation, and radiation response \cite{xia2016phase, lu2016enhancing, granberg2016mechanism, nastasi1991thermodynamics}. Of particular research interest are radiation-induced segregation \cite{nastar20121, thuinet2018multiscale, wiedersich1979theory} and radiation-induced precipitation \cite{jiao2011novel, meslin2013radiation, cauvin1981solid}, which generate inhomogeneities that can negatively affect material response, and are controlled by kinetic parameters such as transport coefficients.

Particularly promising candidates for structural materials are refractory high entropy alloys \cite{senkov2010refractory, senkov2011mechanical, el2019outstanding, el2023quinary, sun2025thermodynamic}. As such, understanding and modeling PD-mediated diffusion in refractory high entropy alloys, and concentrated alloys in general, is necessary for alloy discovery for advanced applications. Modeling PD-mediated diffusion in these systems is nontrivial due to the large variations in local chemical environments seen by the diffusing PD over time \cite{wang2022disentangling, zhao2021role}. The conventional approach to modeling PD-mediated diffusion in these systems is to fit the energy barrier of a single PD hop using the kinetically-resolved action (KRA) expression \cite{PhysRevB.64.184307, PhysRevLett.94.045901, zhang2019kinetically}:

\begin{equation}
    \Delta E = E_\text{KRA} + \frac{E' - E^\circ}{2}
\end{equation}

where $E_\text{KRA}$ is a fitting parameter, which may depend on local chemical environment, $E^\circ$ and $E'$ are respectively the beginning and final energies along the reaction coordinate of the hop, and $\Delta E$ is the energy barrier for the hop. Then, energies are fit as a function of local chemical embeddings, usually with the cluster expansion (CE) technique \cite{van2018first, de1994cluster} and density functional theory (DFT) calculations \cite{hohenberg1964inhomogeneous, kohn1965self} or otherwise atomistic data. Therefore, if the KRA expression is suitably accurate, and we have sufficient first-principles data to fit a KRA model, we can easily run kinetic Monte Carlo (KMC) simulations \cite{bortz1975new} to compute kinetic quantities of interest, e.g. diffusion coefficients and transport coefficients.

A particular success of the KRA expression is that all the quantities within the expression are independent of direction along the transition pathway, making it very well-suited for CE models. Then, the only directionality is encoded within the last term, which can be interpreted as a thermodynamic driving force. Thus, to then recover asymmetric energy barriers, it must be true that $\Delta E$ is at least moderately correlated with $E' - E^\circ$.

In this work, we find that this assumption is not met for vacancy hopping in Ta-W alloys computed using the nudged elastic band method (NEB) method \cite{henkelman2000improved, henkelman2000climbing, nakano2008space, maras2016global} and DFT. We additionally hypothesize that this correlation is weak in a large class of concentrated alloys. As such, it is advantageous to not rely on a model such as the KRA, which assumes such correlations, and instead compute energy barriers directly from some embedding of local chemistry.

Li et al., for example, used a transition-state cluster expansion to predict activation energies for vacancy migration in Pt-Ni nanoparticles, and found that such a description was a better predictor for energy barriers than the KRA relation \cite{li2021predicting}. This was achieved by explicitly including midpoint lattice sites within cluster space, allowing transition states to be explicitly enumerated within cluster space. They found, however, that computing the activation energy in this fashion was very numerically expensive relative to the cheaper broken-bond model, since the energy of both the initial and transition states are evaluated separately to predict the energy barrier of the event. Furthermore, their work was restricted to vacancy migration, and therefore needs to be extended to describe a larger class of diffusive mechanisms.

Beyond just vacancy diffusion, the energy barrier of a thermally activated event should largely depend on local interactions, akin to a broken bond model. In our prior work, we developed a tensorial cluster expansion (TCE) technique that allowed us to truncate away terms that are nonlocal to a swap between two atoms, significantly decreasing the computation time needed to predict the energy difference of said swap. Additionally, said technique is agnostic of the lattice structure of the system in question \cite{jeffries2026generalized}.

In this work, we extend our TCE technique to include transition states, akin to Li et. al \cite{li2021predicting}. Using this extended model, we showcase how one can describe vacancy migration, as well as the migration and rotation of self-interstitial (SIA) atoms in bcc solids, and argue that any thermally activated event on a static lattice can be described by this extension. We then fit this extended model to vacancy migration in Ta-W using first-principles data using a lower-dimensional embedding of the migration data, showcasing an effective clustering of vacancy migration pathways. Lastly, we then use our fitted model to compute diffusivity over a large temperature and composition range, finding four categorically distinct regimes over composition. Notably, we find that dilute Ta inhibits vacancy diffusion up to a percolation threshold, and that dilute W leaves vacancy diffusion largely unaffected.

%% file: sections/model.tex
\section{Model}

First, following a previously developed tensorial cluster expansion model \cite{jeffries2026generalized}, we write our energy as a linear function of the number of $n$'th order $\alpha$-$\beta$ pairs and the number of $n$'th order $\alpha$-$\beta$-$\gamma$ three body clusters:

\begin{equation}
    \begin{aligned}
        E(\mathbf{X}) &= \frac{1}{2!}\varepsilon_{\alpha\beta}^{(n)}N_{\alpha\beta}^{(n)}(\mathbf{X}) + \frac{1}{3!}\zeta_{\alpha\beta\gamma}^{(n)}M_{\alpha\beta\gamma}^{(n)}(\mathbf{X})
    \end{aligned}
\end{equation}

where $N_{\alpha\beta}^{(n)}$ is the number of $\alpha$-$\beta$ pairs of $n$'th order, $M_{\alpha\beta\gamma}^{(n)}$ is the number of $\alpha$-$\beta$-$\gamma$ triplets of $n$'th order, and $\varepsilon_{\alpha\beta}^{(n)}$ and $\zeta_{\alpha\beta\gamma}^{(n)}$ are their respective interaction energies. Here, both cluster count features depend on the configuration $\mathbf{X}$. With respect to the lower indices, all of the above tensors are symmetric, necessitating the overcounting factors $1/2!$ and $1/3!$. This can be more easily cast as a simple linear problem:

\begin{equation}
    E(\mathbf{X}) = \mathbf{j}^\intercal\mathbf{t}(\mathbf{X})
\end{equation}

where $\mathbf{X}$ describes the occupation of each lattice site on a static lattice:

\begin{equation}
    X_{i\alpha} = \begin{cases}
        1 & \text{site $i$ is occupied by atom type $\alpha$} \\
        0 & \text{else}
    \end{cases}
\end{equation}

which allows us to calculate the counts of each cluster type as polynomials in $\mathbf{X}$:

\begin{equation}
    \begin{aligned}
        N_{\alpha\beta}^{(n)}(\mathbf{X}) &= A_{ij}^{(n)}X_{i\alpha}X_{j\beta} \\
        M_{\alpha\beta\gamma}^{(n)}(\mathbf{X}) &= B_{ijk}^{(n)}X_{i\alpha}X_{j\beta}X_{k\gamma}
    \end{aligned}
\end{equation}

where $A_{ij}^{(n)}$ and $B_{ijk}^{(n)}$ are symmetric tensors encoding the topology of the lattice:

\begin{equation}
    \begin{aligned}
        A_{ij}^{(n)} &= \begin{cases}
            1 & (i,j)\in \text{$n$'th nearest neighbors} \\
            0 & \text{else}
        \end{cases} \\
        B_{ijk}^{(n)} &= \begin{cases}
            1 & (i,j,k)\in \text{$n$'th order triplets} \\
            0 & \text{else}
        \end{cases}
    \end{aligned}
\end{equation}

Then, $\mathbf{t}$ is a column vector containing the cluster counts:

\begin{equation}
    \mathbf{t}(\mathbf{X}) = \begin{pmatrix}
        \text{vec}(\mathbf{N}(\mathbf{X}))^\intercal &
        \text{vec}(\mathbf{M}(\mathbf{X}))^\intercal
    \end{pmatrix}^\intercal
\end{equation}

and $\mathbf{j}$ is a learnable column vector of interaction parameters:

\begin{equation}\label{eq:ECIs}
    \mathbf{j} = \begin{pmatrix}
        \displaystyle\frac{1}{2!}\text{vec}(\boldsymbol{\varepsilon})^\intercal &
        \displaystyle\frac{1}{3!}\text{vec}(\boldsymbol{\zeta})^\intercal
    \end{pmatrix}^\intercal
\end{equation}

For a given reaction pathway $\mathbf{X}^\circ \to \mathbf{X}^\ddagger \to \mathbf{X}'$, respectively denoting the initial, transition, and final states, the transition rate can be approximated by harmonic transition state theory (HTST) \cite{eyring1935activated}:

\begin{equation}
    r = \nu \exp\left(-\frac{E^\ddagger - E^\circ}{k_BT}\right)
\end{equation}

where $E^\ddagger$ and $E^\circ$ are respectively the energies of the transition and initial states, and $\nu$ is an attempt frequency. Rather than predict absolute energies, we can instead use the feature vector differences to predict energy differences directly, similarly to Musa et al. \cite{MUSA2025116535}. This naturally fits into the description above, in which we need energy barriers to run KMC, i.e. energy differences between initial and saddle states. We can model the energy barrier using the feature vectors above:

\begin{equation}
    \begin{aligned}
        E^\ddagger - E^\circ &= \mathbf{j}^\intercal\left(\mathbf{t}(\mathbf{X}^\ddagger) - \mathbf{t}(\mathbf{X}^\circ)\right) \\
        &= \mathbf{j}^\intercal\left(\mathbf{t}^\ddagger - \mathbf{t}^\circ\right)
    \end{aligned}
\end{equation}

which is a generalized broken bond model, since the first $k^2$ elements of $\mathbf{t}^\ddagger - \mathbf{t}^\circ = \mathbf{t}(\mathbf{X}^\ddagger) - \mathbf{t}(\mathbf{X}^\circ)$ represent the pairwise interactions at the initial state that are broken at the transition state, where $k$ is the number of chemical species in the system. From our prior work, we can also compute this feature vector difference as a function of local topology \cite{jeffries2026generalized}, significantly reducing the computational cost of predicting the energy barrier. This feature vector difference is:

\begin{equation}\label{eq:feat-vec-diff}
    \begin{aligned}
        \Delta \Tilde{N}_{\alpha\beta}^{(n)} &= \sum_{d\in\mathcal{D}} A_{dj}^{(n)}X_{d\alpha}^\ddagger X_{j\beta}^\ddagger\\
        &- \sum_{d\in\mathcal{D}} A_{dj}^{(n)}X_{d\alpha}^\circ X_{j\beta}^\circ \\
        \Delta N_{\alpha\beta}^{(n)} &= 2\Delta\Tilde{N}_{(\alpha\beta)}^{(n)} \\
        \Delta \Tilde{M}_{\alpha\beta\gamma}^{(n)} &= \sum_{d\in\mathcal{D}} B_{djk}^{(n)}X_{d\alpha}^\ddagger X_{j\beta}^\ddagger X_{k\gamma}^\ddagger\\
        &- \sum_{d\in\mathcal{D}}B_{djk}^{(n)}X_{d\alpha}^\circ X_{j\beta}^\circ X_{k\gamma}^\circ \\
        \Delta M_{\alpha\beta\gamma}^{(n)} &= 3\Delta \Tilde{M}_{(\alpha\beta\gamma)}^{(n)} \\
        \mathbf{t}^\ddagger - \mathbf{t}^\circ &= \begin{pmatrix}
            \text{vec}(\Delta \mathbf{N}) \\
            \text{vec}(\Delta\mathbf{M})
        \end{pmatrix}
    \end{aligned}
\end{equation}

where parenthesis denote symmetrization of the tensor, and $\mathcal{D} = \{i: \mathbf{X}_i^\ddagger \neq \mathbf{X}_i^\circ\}$ is the set of lattice sites that change between the initial and transition states. This model additionally obeys detailed balance, which is necessary for correctly sampling the Boltzmann distribution \cite{voter2007introduction}. Namely, consider a reaction pathway $\mathbf{X}^\circ \to \mathbf{X}^\ddagger \to \mathbf{X}'$ and its corresponding reverse reaction pathway $\mathbf{X}' \to \mathbf{X}^\ddagger \to \mathbf{X}^\circ$. Then:

\begin{equation}
    \begin{aligned}
        p(\mathbf{X}^\circ) r(\mathbf{X}^\circ\to\mathbf{X}') &= \frac{1}{Z}e^{-\beta\mathbf{j}^\intercal\mathbf{t}^\circ} \nu e^{-\beta\mathbf{j}^\intercal\left(\mathbf{t}^\ddagger - \mathbf{t}^\circ\right)} \\
        &=\frac{\nu}{Z}e^{-\beta \mathbf{j}^\intercal\mathbf{t}^\ddagger} \\
        &=\frac{1}{Z}e^{-\beta\mathbf{j}^\intercal\mathbf{t}'}\nu e^{-\beta \mathbf{j}^\intercal\left(\mathbf{t}^\ddagger - \mathbf{t}'\right)} \\
        &=p(\mathbf{X}')r(\mathbf{X}'\to\mathbf{X}^\circ)
    \end{aligned}
\end{equation}

where $p(\mathbf{X})$ denotes the probability of a state, $r(\mathbf{X}\to\mathbf{Y})$ denotes the rate of a transition, and $Z = \sum_{\mathbf{X}^\circ} e^{-\beta\mathbf{j}^\intercal\mathbf{t}(\mathbf{X}^\circ)}$ is the partition function of the system.

In conventional CE-based treatments of diffusion, the CE is typically used to describe the energies of stable initial and final lattice configurations, while the transition state is not treated as an independent configuration with its own cluster representation. This is typically addressed by correlating the transition state energy with the initial and final energies of the transition (i.e. the KRA relation), and using this correlation to approximate the barrier for that transition. This approach assumes that the local energetics of the initial and final states provide a reliable proxy for the energetics of the transition state. In the present work, we find that this correlation is weak at best, motivating a formulation in which the transition state itself is explicitly decorated with a feature vector, rather than its energetics inferred solely from the endpoint configurations.

\begin{figure}[H]
    \centering
    \includegraphics[width=0.9\linewidth]{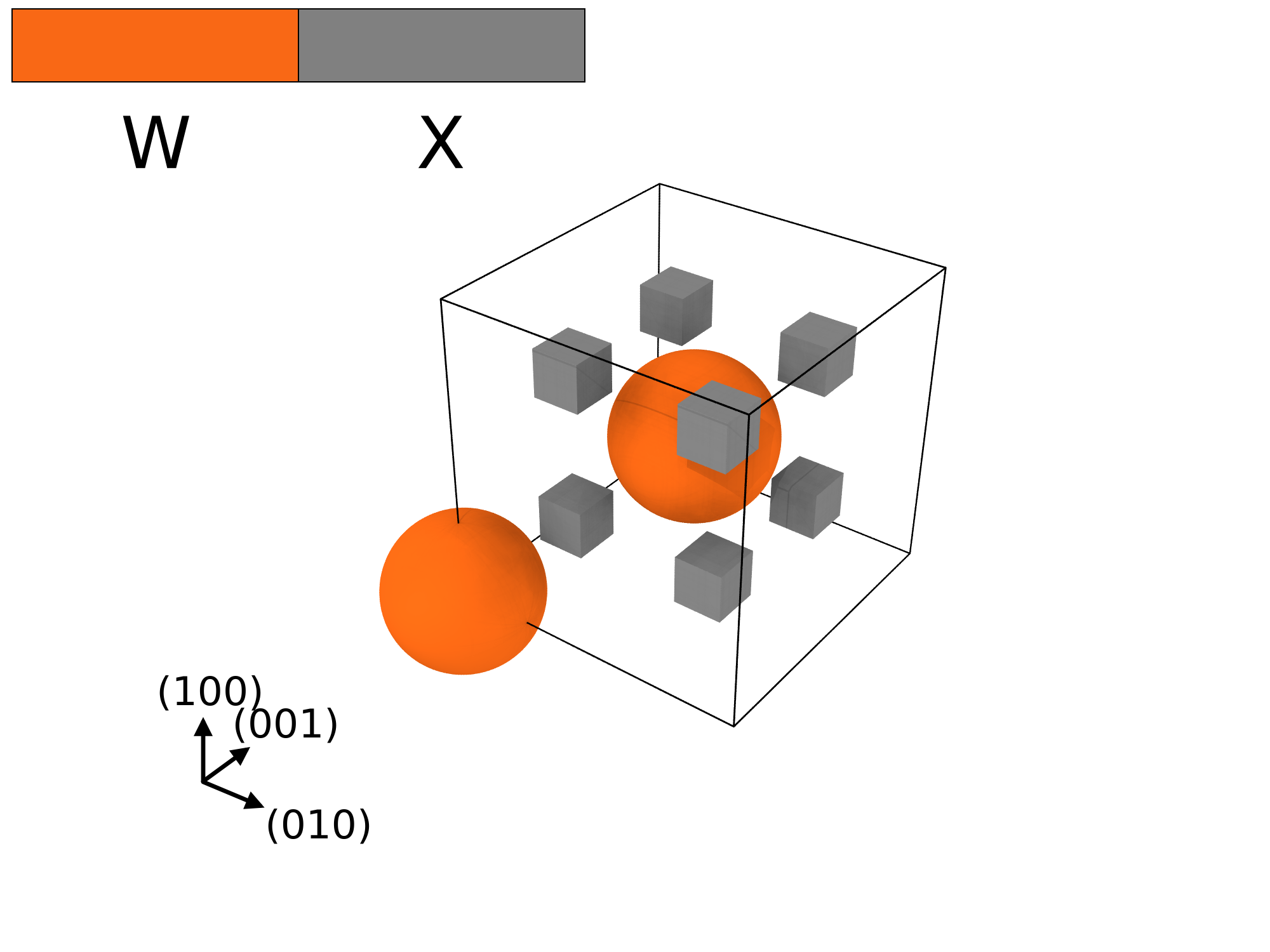}
    \caption{BCC multi-lattice including transition states with the ordinary BCC sites denoted by tungsten (W) and the transition state sites denoted by X, rendered with Open Visualization Tool (OVITO) \cite{ovito}. Topological features, i.e. nearest neighbor distances and three body labels, of this multi-lattice are included in Table~\ref{tab:orders}.}
    \label{fig:bcc-multilattice}
\end{figure}

Li et al. found that by augmenting configuration space to explicitly include transition states, allowing us to decorate the initial, transition, and final states of a vacancy hop within the same feature space, the energy barrier of a vacancy migration could be more accurately predicted than by using the KRA relation\cite{li2021predicting}. These transition state lattice sites sit in-between the standard lattice sites, allowing the hopping atom to occupy space between the standard lattice sites during a vacancy hop. In body-centered cubic (bcc) crystal structures, for example, the unit cell of such a multi-lattice is the ordinary bcc lattice positions ($\{(0, 0, 0), (1/2, 1/2, 1/2)\}$) with transition state lattice positions around the central atom in the directions of its nearest neighbors, i.e. $(1/2, 1/2, 1/2) + \frac{1}{4}(\pm 1, \pm 1, \pm 1)$ (Figure \ref{fig:bcc-multilattice}). Note that this specific augmented configuration space is tailored specifically to first nearest neighbor hops in bcc solids, i.e. vacancy hops along $\langle 111\rangle$ directions. However, one could include additional transition state lattice sites for other hops, e.g. $(0, 0, 1/2)$ points for diffusion along $\langle 100\rangle$ directions in bcc solids. However, in this work, we only account for diffusion along $\langle 111\rangle$ directions, since diffusion along these directions is more favorable than along $\langle 100\rangle$ directions in pure W \cite{oda2014first}.

\begin{table}[H]
    \centering
    \begin{tabular}{|c|c|c|}
        \hline
        order & neighbor distance & three body label \\
        \hline
        $1$ & $\sqrt{3}a/4 \approx 0.43a$ & $(1, 1, 2)$ \\
        \hline
        $2$ & $a/2=0.50a$ & $(1, 1, 3)$ \\
        \hline
        $3$ & $\sqrt{2}a/2\approx 0.71a$ & $(2, 2, 3)$ \\
        \hline
        $4$ & $\sqrt{11}a/4 \approx 0.83a$ & $(3, 3, 3)$ \\
        \hline
        $5$ & $\sqrt{3}a/2 \approx 0.87a$ & $(1, 2, 4)$ \\
        \hline
        $6$ & $a = 1.00a$ & $(1, 3, 4)$ \\
        \hline
        $7$ & $\sqrt{19}a/4 \approx 1.09a$ & $(1, 4, 4)$ \\
        \hline
        $8$ & $\sqrt{5}a/2 \approx 1.12a$ & $(2, 4, 4)$ \\
        \hline
    \end{tabular}
    \caption{Neighbor distances and three body labels in the multi-lattice, up to order $8$.}
    \label{tab:orders}
\end{table}

However, Li et al.'s framework explicitly evaluates the energy of both the initial and transition states, making the evaluation of energy barriers, and therefore any simulation involving the evaluation of energy barriers, significantly slower than a broken bond model \cite{li2021predicting}. Here, because the energy barrier depends explicitly on the count differences $\Delta\mathbf{N}$ and $\Delta\mathbf{M}$, we can efficiently compute energy barriers as a local difference in features, akin to a broken bond model, rather than globally recomputing redundant interactions far away from the activated event.

\begin{figure}[H]
    \centering
    \includegraphics[width=\linewidth]{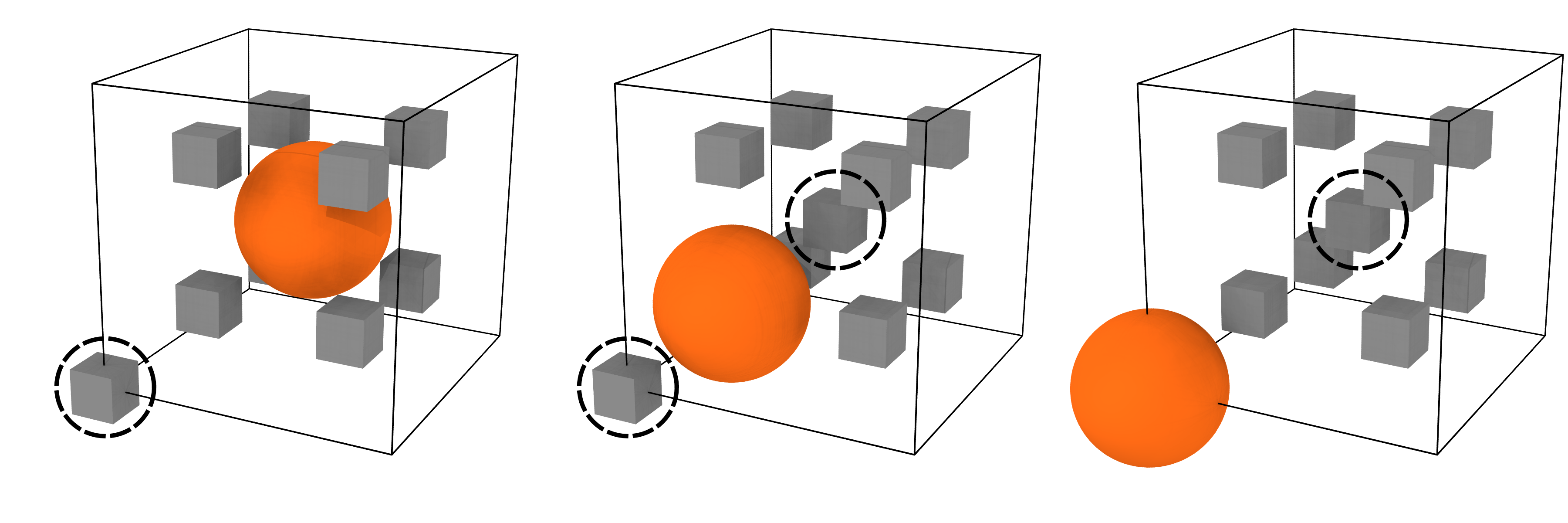}
    \definecolor{tiger-orange}{HTML}{F96815}
    \begin{tikzpicture}[
        x=2.2cm,
        y=1.4cm,
        atom/.style={circle, draw=black, fill=tiger-orange, minimum size=6.5mm, inner sep=0pt},
        vac/.style={rectangle, draw=black, fill=gray!55, minimum size=5.5mm, inner sep=0pt},
        lbl/.style={font=\small},
        steplbl/.style={font=\small},
        bondlbl/.style={font=\scriptsize}
    ]
    
        \def\xA{0}
        \def\xB{1}
        \def\xC{2}
        
        \def\yOne{0}
        \def\yTwo{-1.2}
        \def\yThree{-2.4}
        
        \node[lbl] at (\xA,\yOne+0.45) {$A^\circ$};
        \node[lbl] at (\xB,\yOne+0.45) {$A^\ddagger$};
        \node[lbl] at (\xC,\yOne+0.45) {$A'$};
        
        \draw[->, thick] (\xA-0.1,\yThree-1.2) -- (\xC+0.1,\yThree-1.2)
            node[midway, above=2pt] {$[111]$};
        
        \node[steplbl, anchor=east] at (\xA-0.45,\yOne) {initial};
        \node[steplbl, anchor=east] at (\xA-0.45,\yTwo) {transition};
        \node[steplbl, anchor=east] at (\xA-0.45,\yThree) {final};
        
        \draw[thin, black!25] (\xA-0.05,\yOne) -- (\xC+0.05,\yOne);
        \draw[thin, black!25] (\xA-0.05,\yTwo) -- (\xC+0.05,\yTwo);
        \draw[thin, black!25] (\xA-0.05,\yThree) -- (\xC+0.05,\yThree);
        
        \draw[dashed] (\xA,\yOne) -- (\xB,\yOne);
        \draw[dashed] (\xB,\yOne) -- (\xC,\yOne);
        \draw[dashed] (\xA,\yOne) to[bend right=40] (\xC,\yOne);
        
        \node[bondlbl] at (0.5,\yOne+0.2) {$\varepsilon_{\text{X}\text{X}}^{(1)}$};
        \node[bondlbl] at (1.5,\yOne+0.2) {$\varepsilon_{\text{W}\text{X}}^{(1)}$};
        \node[bondlbl] at (1.0,\yOne-0.4) {$\varepsilon_{\text{W}\text{X}}^{(5)}$};
        
        \node[vac]  at (\xA,\yOne) {};
        \node[vac]  at (\xB,\yOne) {};
        \node[atom] at (\xC,\yOne) {};
        
        \draw[dashed] (\xA,\yTwo) -- (\xB,\yTwo);
        \draw[dashed] (\xB,\yTwo) -- (\xC,\yTwo);
        \draw[dashed] (\xA,\yTwo) to[bend right=40] (\xC,\yTwo);
        
        \node[bondlbl] at (0.5,\yTwo+0.2) {$\varepsilon_{\text{W}\text{X}}^{(1)}$};
        \node[bondlbl] at (1.5,\yTwo+0.2) {$\varepsilon_{\text{W}\text{X}}^{(1)}$};
        \node[bondlbl] at (1.0,\yTwo-0.4) {$\varepsilon_{\text{X}\text{X}}^{(5)}$};
        
        \node[vac]  at (\xA,\yTwo) {};
        \node[atom] at (\xB,\yTwo) {};
        \node[vac]  at (\xC,\yTwo) {};
        
        \draw[dashed] (\xA,\yThree) -- (\xB,\yThree);
        \draw[dashed] (\xB,\yThree) -- (\xC,\yThree);
        \draw[dashed] (\xA,\yThree) to[bend right=40] (\xC,\yThree);
        
        \node[bondlbl] at (0.5,\yThree+0.2) {$\varepsilon_{\text{W}\text{X}}^{(1)}$};
        \node[bondlbl] at (1.5,\yThree+0.2) {$\varepsilon_{\text{X}\text{X}}^{(1)}$};
        \node[bondlbl] at (1.0,\yThree-0.4) {$\varepsilon_{\text{W}\text{X}}^{(5)}$};
        
        \node[atom] at (\xA,\yThree) {};
        \node[vac]  at (\xB,\yThree) {};
        \node[vac]  at (\xC,\yThree) {};
    
    \end{tikzpicture}
    \caption{Vacancy migration on the bcc multi-lattice with the initial, transition, and final states with the migrating vacancy circled, with delocalization of the migrating vacancy in the transition state (top) and cartoon of migration with bond energies labeled (bottom).}
    \label{fig:bcc-pathway}
\end{figure}
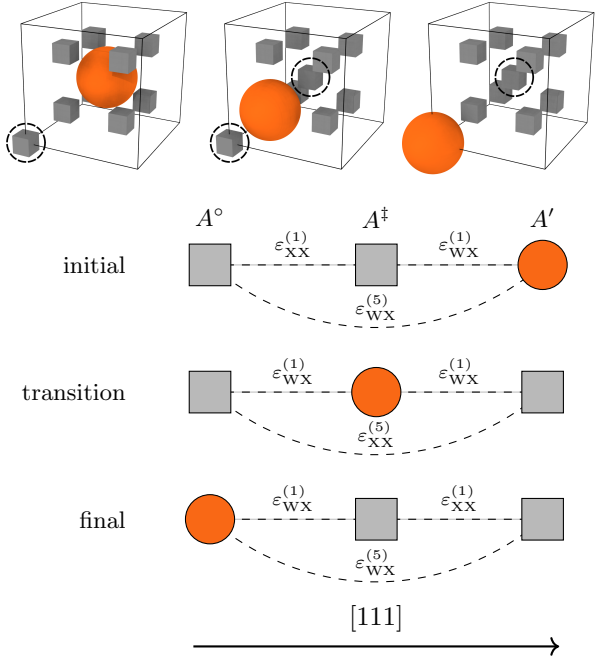

From this description, we can model migration pathways as static configurations on a multi-lattice. For example, for a vacancy migration in a bcc system, we can place a vacancy at the $(0, 0, 0)$ site and an atom at the $(1/2, 1/2, 1/2)$ site. The corresponding transition state is then ``semi-vacancies" at the $(0, 0, 0)$ and $(1/2, 1/2, 1/2)$ sites and an atom at the $(1/4, 1/4, 1/4)$ midpoint (Fig.~\ref{fig:bcc-pathway}).

Within this multi-lattice, we explicitly include the X type as an atomic type within the enumerated clusters. This is convenient for decorating transition states with feature vectors, but also allows the training to encode physical features relating to vacancies, e.g. vacancy-vacancy binding energies.

Then, for a given initial state with a vacancy at site $A^\circ$ and configuration tensor $\mathbf{X}^\circ$, we can construct the transition and final states, respectively denoted $\mathbf{X}^\ddagger$ and $\mathbf{X}'$, for a vacancy migration $\mathbf{X}^\circ\to\mathbf{X}'$:

\begin{equation}
    \begin{aligned}
        \mathbf{X}^\ddagger &= \pi_1\mathbf{X}^\circ \\
        \mathbf{X}' &= (\pi_2\circ\pi_1)\mathbf{X}^\circ
    \end{aligned}
\end{equation}

where $\pi_1, \pi_2\in S_n$ are permutations that transform the configurations by acting on the positional indices of the tensors, $S_n$ is the symmetric group of order $n$, and $n$ is the number of lattice sites within the supercell constructed with the multi-lattice. We have adopted the convention that the right-most permutation acts first. Here, permutations $\pi\in S_n$ act on configurational tensors $\mathbf{X}$ by means of a linear transformation:

\begin{equation}
    (\pi\mathbf{X})_{i\alpha} = \delta_{i\pi(j)}X_{j\alpha} = X_{\pi^{-1}(i)\alpha}
\end{equation}

Then, written in cycle notation, the permutations above are:

\begin{equation}
    \begin{aligned}
        \pi_1 &= (i^\ddagger i') \\
        \pi_2 &= (i^\circ i^\ddagger) \\
        \pi_2 \circ \pi_1 &= (i^\circ i^\ddagger)\circ (i^\ddagger i') = (i^\circ i')
    \end{aligned}
\end{equation}

where $i^\circ$, $i^\ddagger$, and $i'$ are respectively the indices of lattice sites $A^\circ$, $A^\ddagger$, and $A'$, which are relative to the position of the migrating vacancy. This notation allows us to represent arbitrarily complex defect migration mechanisms on this fixed lattice within the canonical ensemble. Namely, the number of $\alpha$ types within the system is invariant under these permutations since any such $\pi$ is a bijection. Additionally, the set of lattice sites that change upon this permutation is then exactly $\text{supp}(\pi)$, i.e. the support of the permutation, which can be plugged into Eq.~\eqref{eq:feat-vec-diff} to efficiently compute feature vector differences.

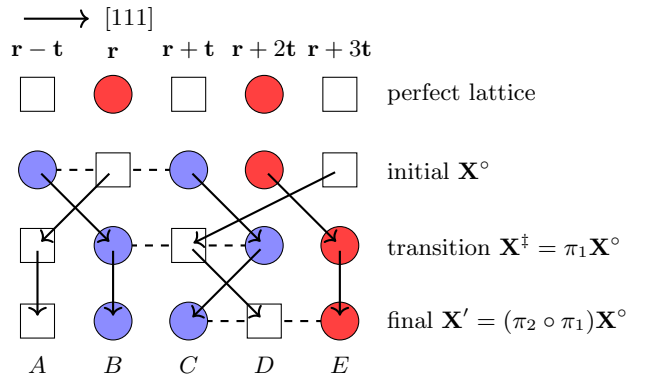
\begin{figure}[H]
    \centering
    \begin{tikzpicture}[
        x=1.0cm,
        y=1.0cm,
        atom/.style={circle, draw=black, fill=red!75, minimum size=5mm, inner sep=0pt},
        migrating/.style={circle, draw=black, fill=blue!45, minimum size=5mm, inner sep=0pt},
        ghost/.style={rectangle, draw=black, fill=white, minimum size=4.5mm, inner sep=0pt},
        arr/.style={->, thick, shorten >=2pt, shorten <=2pt},
        darr/.style={dashed, thick},
        lbl/.style={font=\small},
        note/.style={font=\small}
    ]
    
    \coordinate (A0) at (0, 2);
    \coordinate (A1) at (0, 1);
    \coordinate (A2) at (0, 0);
    \coordinate (A3) at (0, -1);
    
    \coordinate (B0) at (1, 2);
    \coordinate (B1) at (1, 1);
    \coordinate (B2) at (1, 0);
    \coordinate (B3) at (1, -1);
    
    \coordinate (C0) at (2, 2);
    \coordinate (C1) at (2, 1);
    \coordinate (C2) at (2, 0);
    \coordinate (C3) at (2, -1);
    
    \coordinate (D0) at (3, 2);
    \coordinate (D1) at (3, 1);
    \coordinate (D2) at (3, 0);
    \coordinate (D3) at (3, -1);
    
    \coordinate (E0) at (4, 2);
    \coordinate (E1) at (4, 1);
    \coordinate (E2) at (4, 0);
    \coordinate (E3) at (4, -1);
    
    \node[ghost]  at (A0) {};
    \node[atom] at (B0) {};
    \node[ghost]  at (C0) {};
    \node[atom] at (D0) {};
    \node[ghost]  at (E0) {};
    
    \node[note, right=0.5cm of E0] {perfect lattice};
    
    
    \draw[darr] (A1) -- (C1);
    \draw[darr] (B2) -- (D2);
    \draw[darr] (C3) -- (E3);
    
    \node[migrating]  at (A1) {};
    \node[ghost] at (B1) {};
    \node[migrating]  at (C1) {};
    \node[atom] at (D1) {};
    \node[ghost]  at (E1) {};
    
    \node[note, right=0.5cm of E1] {initial $\mathbf{X}^\circ$};
    
    \node[ghost]  at (A2) {};
    \node[migrating]   at (B2) {};
    \node[ghost] at (C2) {};
    \node[migrating]   at (D2) {};
    \node[atom] at (E2) {};
    
    \node[note, right=0.5cm of E2] {transition $\mathbf{X}^\ddagger = \pi_1\mathbf{X}^\circ$};
    
    \node[ghost]  at (A3) {};
    \node[migrating]   at (B3) {};
    \node[migrating] at (C3) {};
    \node[ghost]   at (D3) {};
    \node[atom] at (E3) {};
    
    \node[note, right=0.5cm of E3] {final $\mathbf{X}' = (\pi_2\circ\pi_1)\mathbf{X}^\circ$};
    
    \draw[arr] (A1) -- (B2);
    \draw[arr] (B1) -- (A2);
    \draw[arr] (C1) -- (D2);
    \draw[arr] (D1) -- (E2);
    \draw[arr] (E1) -- (C2);
    
    \draw[arr] (A2) -- (A3);
    \draw[arr] (B2) -- (B3);
    \draw[arr] (C2) -- (D3);
    \draw[arr] (D2) -- (C3);
    \draw[arr] (E2) -- (E3);
    
    \node[lbl, below=10pt of A3] {$A$};
    \node[lbl, below=10pt of B3] {$B$};
    \node[lbl, below=10pt of C3] {$C$};
    \node[lbl, below=10pt of D3] {$D$};
    \node[lbl, below=10pt of E3] {$E$};
    
    \node[lbl, above=10pt of A0] {$\mathbf r - \mathbf t$};
    \node[lbl, above=10pt of B0] {$\mathbf r$};
    \node[lbl, above=10pt of C0] {$\mathbf r + \mathbf{t}$};
    \node[lbl, above=10pt of D0] {$\mathbf r + 2\mathbf t$};
    \node[lbl, above=10pt of E0] {$\mathbf r + 3\mathbf t$};
    
    \draw[arr] (-0.25,3) -- (0.75,3)
        node[right,lbl] {$[111]$};
    
    \end{tikzpicture}
    \caption{SIA migration along the $[111]$ direction via the permutation $\pi_1 = (AB)(CDE)$ and $\pi_2 = (CD)$. Here, the SIA center migrates from point $B$, at $\mathbf{r}\in\mathbb{R}^3$, to point $D$, at $\mathbf{r} + 2\mathbf{t} \in \mathbb{R}^3$, where $\mathbf{t}$ is the translation vector along the $[111]$ direction. Here, $\mathbf{t}$ is parallel to $(1, 1, 1)$, and has magnitude $d_{\langle 111\rangle}/2$, where $d_{\langle 111\rangle}$ is the inter-planar distance along the $\langle 111\rangle$ direction in bcc.}
    \label{fig:sia-migration}
\end{figure}

In general, within the canonical ensemble, i.e. where the concentration of the system is preserved from an event $\mathbf{X}\to\mathbf{X}'$, we can describe any event via a permutation $\pi$ acting on $\mathbf{X}$. For example, we can describe the migration of an SIA in a bcc solid along the $[111]$ direction using the same multi-lattice, i.e. with midpoints in between lattice sites (Fig.~\ref{fig:sia-migration}).

\begin{figure}[H]
    \centering
    \begin{tikzpicture}[
        x=0.35cm,
        y=0.35cm,
        atom/.style={circle, draw=black, fill=red!75, minimum size=3mm, inner sep=0pt},
        migrating/.style={circle, draw=black, fill=blue!45, minimum size=3mm, inner sep=0pt},
        ghost/.style={rectangle, draw=black, fill=white, minimum size=3mm, inner sep=0pt},
        arr/.style={->, thick, shorten >=2pt, shorten <=2pt},
        darr/.style={dashed, thick},
        lbl/.style={font=\small},
        note/.style={font=\small}
    ]
    
    \coordinate (A0) at (0, 0);
    \coordinate (B0) at (0, 2);
    \coordinate (C0) at (1.414, 1.414);
    \coordinate (D0) at (2, 0);
    \coordinate (E0) at (-2, 0);
    \coordinate (F0) at (-1.414, -1.414);
    \coordinate (G0) at (0, -2);
    \node[atom] at (A0) {};
    \node[lbl] at ($(A0)+(12pt,-12pt)$) {$A$};
    \node[ghost] at (B0) {};
    \node[lbl] at ($(B0)+(0,10pt)$) {$B$};
    \node[ghost] at (C0) {};
    \node[lbl] at ($(C0)+(8pt,8pt)$) {$C$};
    \node[ghost] at (D0) {};
    \node[lbl] at ($(D0)+(10pt,0)$) {$D$};
    \node[ghost] at (E0) {};
    \node[lbl] at ($(E0)+(-10pt,0)$) {$E$};
    \node[ghost] at (F0) {};
    \node[lbl] at ($(F0)+(-8pt,-8pt)$) {$F$};
    \node[ghost] at (G0) {};
    \node[lbl] at ($(G0)+(0,-10pt)$) {$G$};
    
    \node[note, below=32.5pt of A0] {perfect lattice};

    \draw[arr] ($(A0)+(12pt,-12pt)$) -- ($(A0)+(2pt,-2pt)$);
    
    \begin{scope}[shift={(0, -8)}]
    \coordinate (A1) at (0, 0);
    \coordinate (B1) at (0, 2);
    \coordinate (C1) at (1.414, 1.414);
    \coordinate (D1) at (2, 0);
    \coordinate (E1) at (-2, 0);
    \coordinate (F1) at (-1.414, -1.414);
    \coordinate (G1) at (0, -2);
    \draw[darr] (B1) -- (G1);
    \node[ghost] at (A1) {};
    \node[migrating] at (B1) {};
    \node[ghost] at (C1) {};
    \node[ghost] at (D1) {};
    \node[ghost] at (E1) {};
    \node[ghost] at (F1) {};
    \node[migrating] at (G1) {};
    
    \node[note, below=35pt of A1] {initial};
    \end{scope}
    
    \begin{scope}[shift={(7,-8)}]
    \coordinate (A2) at (0, 0);
    \coordinate (B2) at (0, 2);
    \coordinate (C2) at (1.414, 1.414);
    \coordinate (D2) at (2, 0);
    \coordinate (E2) at (-2, 0);
    \coordinate (F2) at (-1.414, -1.414);
    \coordinate (G2) at (0, -2);
    \draw[darr] (C2) -- (F2);
    \node[ghost] at (A2) {};
    \node[ghost] at (B2) {};
    \node[migrating] at (C2) {};
    \node[ghost] at (D2) {};
    \node[ghost] at (E2) {};
    \node[migrating] at (F2) {};
    \node[ghost] at (G2) {};
    
    \node[note, below=35pt of A2] {transition};
    \end{scope}
    
    \begin{scope}[shift={(14,-8)}]
    \coordinate (A3) at (0, 0);
    \coordinate (B3) at (0, 2);
    \coordinate (C3) at (1.414, 1.414);
    \coordinate (D3) at (2, 0);
    \coordinate (E3) at (-2, 0);
    \coordinate (F3) at (-1.414, -1.414);
    \coordinate (G3) at (0, -2);
    \draw[darr] (E3) -- (D3);
    \node[ghost] at (A3) {};
    \node[ghost] at (B3) {};
    \node[ghost] at (C3) {};
    \node[migrating] at (D3) {};
    \node[migrating] at (E3) {};
    \node[ghost] at (F3) {};
    \node[ghost] at (G3) {};
    
    \node[note, below=35pt of A3] {final};
    \end{scope}
    
    \draw[arr] ($(A1) + (15pt, -42.5pt)$) -- ($(A2) + (-25pt, -42.5pt)$);
    \node[lbl] at (10.5, -2.5) {$\pi_1 = (BC)(FG)$};
    
    \draw[arr] ($(A2) + (25pt, -42.5pt)$) -- ($(A3) + (-15pt, -42.5pt)$);
    \node[lbl] at (10.5, -3.5) {$\pi_2 = (CD)(EF)$};
    
    \node[lbl] at (10.5, -4.5) {$\pi_2\circ\pi_1 = (BD)(GE)$};
    
    \node[lbl] at ($(A1) + (0, -30pt)$) {$\mathbf{X}^\circ$};
    \node[lbl] at ($(A2) + (0, -30pt)$) {$\mathbf{X}^\ddagger = \pi_1\mathbf{X}^\circ$};
    \node[lbl] at ($(A3) + (0, -30pt)$) {$\mathbf{X}' = (\pi_2\circ\pi_1)\mathbf{X}^\circ$};
    
    \begin{scope}[shift={(10.5, 0)}]
    \draw[arr] (0, 0) -- (0, 2.25)
        node[right,lbl] {$[111]$};
    \draw[arr] (0, 0) -- (2.25, 0)
        node[right,lbl] {$[11\overline{1}]$};
    \draw[arr] (0, 0) -- (1.414, 1.414)
        node[right,lbl] {$[110]$};
    \node at (0, 0) {$\odot$}
        node[left,lbl] {$[\overline{1}10]$};
    \end{scope}
    
    \end{tikzpicture}
    \caption{$[111]\to[11\overline{1}]$ rotation of an SIA centered at point $A$, with transition state $[110]$, in a bcc solid represented on a multi-lattice with permutations $\pi_1$ and $\pi_2$.}
    \label{fig:sia-rotation}
\end{figure}
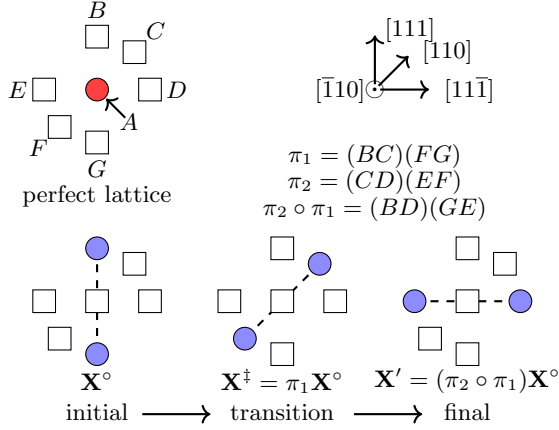

Using more transition state sites, we can also describe more complex dynamics of the SIA, and of random walkers in general. For example, we can describe the $[111]\to[11\overline{1}]$ rotation of an SIA in a bcc solid by permuting on a multi-lattice that also includes $\langle 111\rangle$ and $\langle 110\rangle$ midpoints, assuming that the rotation is a single-barrier event, with the transition state of a $\langle 110\rangle$ SIA (Fig.~\ref{fig:sia-rotation}).

In this work, we strictly focus on vacancy diffusion in Ta-W as the simplest use case. This is due to the potential complexity, e.g. for an SIA rotation, it's not strictly true that a $\langle 111\rangle\to\langle 111\rangle$ SIA is a single-step process, e.g. if the $\langle 110\rangle$ SIAs are stable or metastable. In this case, the local transition graph for rotations is better described by $\langle 111\rangle\to\langle 110\rangle$ transitions, which have low symmetry midpoints. The framework described here is still applicable to such a case, but with low-symmetry midpoints between $\langle 111\rangle$ and $\langle 110\rangle$ states. As such, cases are both extremely system-dependent and nontrivial book-keeping wise. We therefore save numerical explorations of this for a future work.

%% file: sections/methods.tex
\section{Methods and Results}

Using a previously developed CE model \cite{alvarado2023predicting}, we generate structures of varying compositions, namely between $20\%$ and $80\%$ Ta with a $5\%$ step. Each of these structures is then equilibrated using canonical MC implemented in within the Alloy Theoretical Automated Toolkit (ATAT) \cite{van2002alloy} at varying temperatures, namely between $\SI{300}{K}$ and $\SI{1500}{K}$ with a $\SI{300}{K}$ step. This equilibration is merely a sampling technique, i.e. ensuring that our training set includes chemical ordering rather than solely pure samples, and could be easily replaced by a different sampling technique.

\begin{figure}[H]
    \centering
    \includegraphics[width=\linewidth]{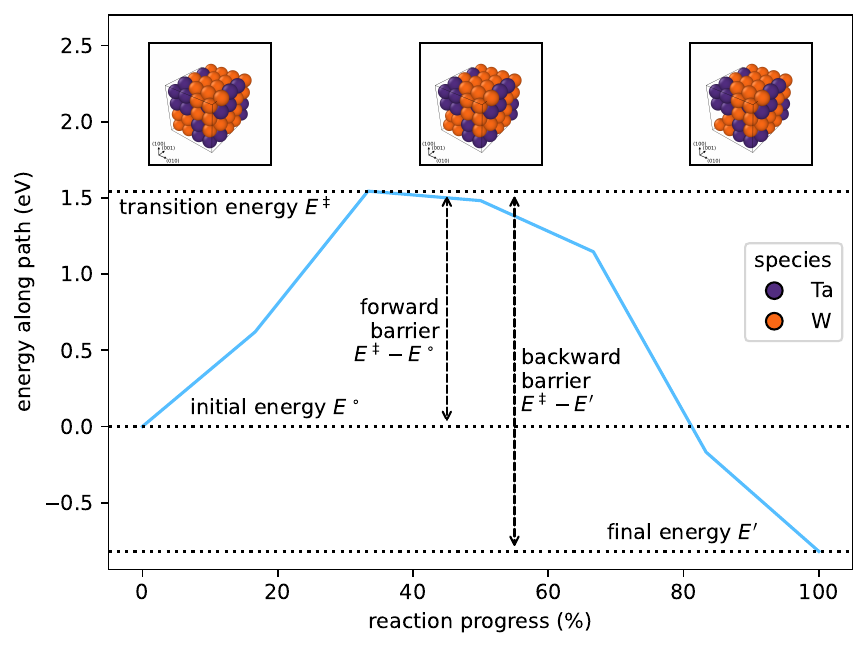}
    \caption{Example minimum energy path computed from NEB and VASP, with the initial, transition, and final states.}
    \label{fig:vasp-neb}
\end{figure}

For each resulting equilibrated structure, we randomly insert a vacancy to generate an initial configuration, and move the vacancy to a random nearest neighbor to generate a final structure. To find the minimum energy path for the hop, we run NEB calculations within the Vienna Ab initio Simulation Package (VASP) \cite{kresse1993ab, kresse1996efficient} with 5 intermediate images. Each image is optimized using the projector augmented-wave method with Perdew-Burke-Ernzerhof exchange correlation functionals \cite{perdew1996generalized} with Gaussian smearing with a width of $\SI{0.1}{eV}$ and energy cutoff of $\SI{500}{eV}$. Electronic self-consistency was achieved with a convergence tolerance of $\SI{1e-5}{eV}$, with a minimum of $5$ and a maximum of $100$ electronic steps per ionic iteration. Forces were converged with a tolerance of $\SI{1e-3}{eV/\AA}$. Each image has fixed cell shape and volume, i.e. only ionic positions were relaxed. An example of one of these minimum energy paths is included in Figure~\ref{fig:vasp-neb}, which generates two data points to train a TCE model, namely the forward and backward hops.

This example trajectory, like many other of the trajectories in the training set, exhibits strong asymmetry, i.e. the transition state lies non-negligibly far away from the midpoint of the hop. From our dataset, we additionally find that $E^\ddagger$ is quite well correlated with $(E'+E^\circ)/2$, with an $r^2$ of nearly $1$ and a slope of nearly $1$. Namely, the KRA approximation for $E^\ddagger$:

\begin{equation}
    E^\ddagger\approx E_\text{KRA} + \frac{E' + E^\circ}{2}
\end{equation}

seems to be valid for our dataset. This results in the expression for the energy barrier $\Delta E$:

\begin{equation}\label{eq:kra-barrier}
    \Delta E = E^\ddagger - E^\circ \approx E_\text{KRA} + \frac{E' - E^\circ}{2}
\end{equation}

\begin{figure}[H]
    \centering
    \includegraphics[width=\linewidth]{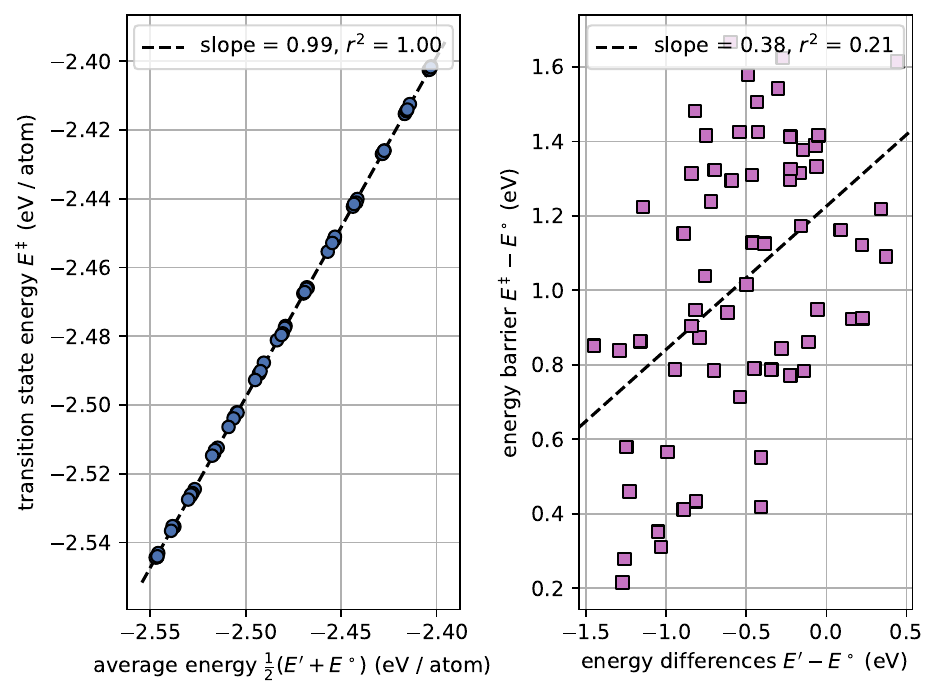}
    \caption{Transition energy $E^\ddagger$ as a function of average energy $(E'+E^\circ)/2$ (left), and energy barrier $E^\ddagger - E^\circ$ as a function of energy difference $E' - E^\circ$ (right) from DFT + NEB calculations of vacancy hopping in Ta-W.}
    \label{fig:kra}
\end{figure}

However, $E_\text{KRA}$ is famously independent of direction, i.e. it is the same for both the forward reaction and its corresponding backwards reaction, so reaction asymmetry is entirely encoded in $E' - E^\circ$ within the KRA energy barrier expression in Eq.~\eqref{eq:kra-barrier}. Therefore, to accurately encode reaction asymmetry, it must be true that $\Delta E$ is at least moderately correlated with $E' - E^\circ$. We find that this correlation is, at best, very weak within our dataset, with a very small $r^2$ of $0.21$ (Fig.~\ref{fig:kra}).

This means that, at least for our dataset, the energy barrier cannot be written as a function of solely $E' - E^\circ$, regardless of how those energies are computed, therefore the KRA barrier expression in Eq.~\eqref{eq:kra-barrier} is insufficient for running vacancy hopping KMC for the Ta-W system. Xi et al. found a similar effect in vacancy migration in the Al-Mg-Zn alloy, in which large fluctuations of the energy barrier were largely explained by variations in $E_\text{KRA}$, rather than $E' - E^\circ$ \cite{xi2022mechanism}, concluding that the standard formulation of the KRA model is insufficient for KMC simulations of vacancy diffusion in that alloy.

For our dataset, we hypothesize that this is a result of the aforementioned strong reaction pathway asymmetry within some of our data points, since the KRA relation assumes that the transition state occurs at the midpoint of the reaction. However, such a hypothesis is difficult to test, since this requires a reinterpretation of the KRA energy $E_\text{KRA}$. As such, we save this testing for a future work. Then, rather than rely on the KRA approximation to compute energy barriers, which will quite poorly approximate energy barriers within our dataset, we will instead compute feature vectors for the initial, transition, and final states within our aforementioned multi-lattice model.

For each minimum-energy path generated by NEB above, indexed by $i$, we then compute the feature vectors of the initial, transition, and final configurations, respectively $\mathbf{t}^\circ_i$, $\mathbf{t}^\ddagger_i$, and $\mathbf{t}'_i$, with energies $E^\circ_i$, $E^\ddagger_i$, and $E'_i$. To compute feature vectors within the multi-lattice, we choose a maximum neighbor order of $8$, and a maximum triplet order of $6$, using a lattice parameter $a = \SI{3.1906}{\AA}$. Note that these neighbor orders are not equivalent to nearest neighbor shells on the ordinary bcc lattice, i.e. that our cluster expansion model does not sample what one would ordinarily consider to be long-range interactions. Within our multi-lattice, an $8$'th nearest neighbor distance corresponds to a distance of $\sqrt{5}a/2 \approx 1.11a$. From these feature vectors, we have two training points for each path, namely:

\begin{equation}
    \begin{aligned}
        p_1 &= (\Delta_\text{f}\mathbf{t}_i, \Delta_\text{f} E_i) \\
        p_2 &= (\Delta_\text{b}\mathbf{t}_i, \Delta_\text{b} E_i)
    \end{aligned}
\end{equation}

where $\Delta_\text{f}\mathbf{t}_i = \mathbf{t}^\ddagger_i - \mathbf{t}^\circ_i$ and $\Delta_\text{f} E_i = E^\ddagger_i - E^\circ_i$ are respectively the feature vector difference and energy barrier for the forward reaction, and $\Delta_\text{b}\mathbf{t}_i = \mathbf{t}^\ddagger_i - \mathbf{t}'_i$ and $\Delta_\text{b}E_i = E^\ddagger_i - E^\circ_i$ are respectively the feature vector difference and energy barrier for the corresponding backward reaction. We note that this lattice parameter depends on composition, so the relaxed configurations from NEB do not exactly match our chosen lattice parameter. We address this by rescaling the supercell to have a lattice parameter equal to that of which we chose prior to computing feature vectors. Using these points, we then seek to fit a linear model using ridge regression:

\begin{equation}\label{eq:loss}
    \begin{aligned}
        L_\text{f}(\mathbf{j}) &= \sum_i \left(\Delta_\text{f}\mathbf{t}_i^\intercal\mathbf{j} - \Delta_\text{f}E_i\right)^2 \\
        L_\text{b}(\mathbf{j}) &= \sum_i\left(\Delta_\text{b}\mathbf{t}_i^\intercal\mathbf{j} - \Delta_\text{b}E_i\right)^2 \\
        L(\mathbf{j}) &= L_\text{f}(\mathbf{j}) + L_\text{b}(\mathbf{j}) + \lambda \|\mathbf{j}\|_2^2\\
        &=\left\|\Delta \mathbf{T}\mathbf{j} - \Delta\mathbf{e}\right\|_2^2 + \lambda \|\mathbf{j}\|_2^2\\
        \mathbf{j}_\text{opt} &= \arg\min_{\mathbf{j}\in\mathbb{R}^k}L(\mathbf{j})\\
        &=\left(\Delta\mathbf{T}^\intercal\Delta\mathbf{T} + \lambda\mathbf{I}\right)^{-1}\Delta\mathbf{T}^\intercal\Delta\mathbf{e}
    \end{aligned}
\end{equation}

where $L_\text{f}$ is the loss for the forward reactions, $L_\text{b}$ is the loss for the backwards reactions, $L$ is the total loss, $\lambda > 0$ is the regularization strength, $\Delta\mathbf{T}$ is the design matrix containing the feature vector differences $\Delta_{\text{f}}\mathbf{t}_i$ and $\Delta_{\text{b}}\mathbf{t}_i$ defined above, $\Delta\mathbf{e}$ is a target vector containing the energy barriers $\Delta_\text{f}E_i$ and $\Delta_\text{b}E_i$ also defined above, $\mathbf{j}$ is the learnable matrix containing cluster interaction energies from Eq.~\eqref{eq:ECIs}, $k = 8\cdot 3^2 + 6\cdot 3^3 = 234$ is the feature size, i.e. the number of chosen clusters, and $\|\cdot\|_2$ denotes the $L_2$ norm. Note that $k$ is not the number of linearly independent clusters, since clusters are both permutation-symmetric and subject to a one-hot occupation constraint $\sum_\alpha X_{i\alpha} = 1$. For our dataset, this count is $\text{rank}(\Delta\mathbf{T}) = 56$.

Using \verb|scikit-learn| \cite{scikit-learn}, we first split our dataset into a training and testing set using an $80\%$/$20\%$ split. Then, we pick the regularization parameter $\lambda$ using leave-one-out cross-validation on the training set, choosing the $\lambda$ that optimizes the resulting root-mean-squared-error (RMSE), resulting in $\lambda = 23.1$.

\begin{figure}[H]
    \centering
    \includegraphics[width=\linewidth]{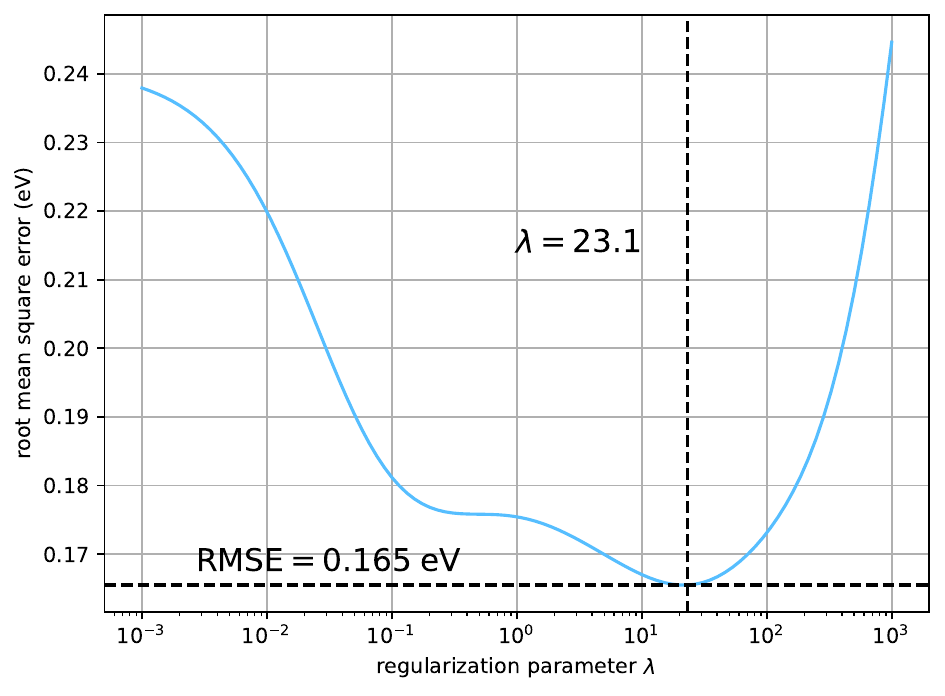}
    \caption{RMSE as a function of regularization parameter using leave-one-out cross validation.}
    \label{fig:cross-val}
\end{figure}

\begin{figure}[H]
    \centering
    \includegraphics[width=\linewidth]{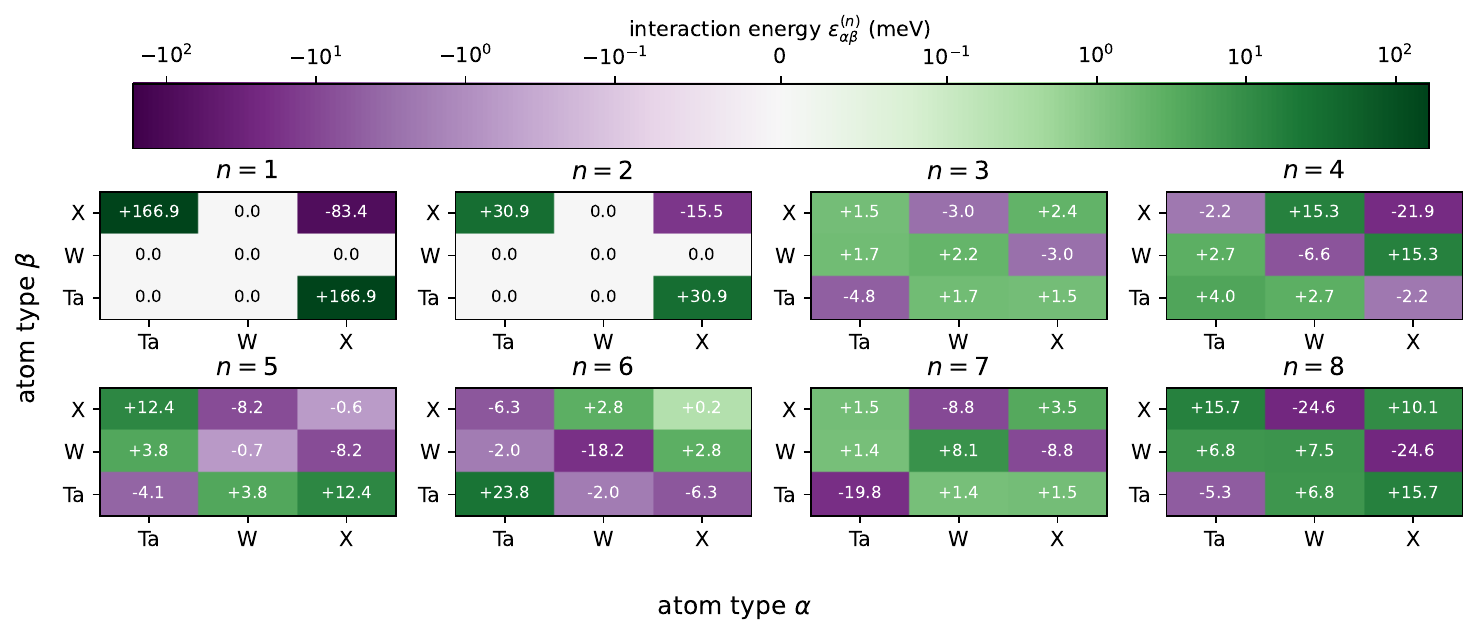}
    \includegraphics[width=\linewidth]{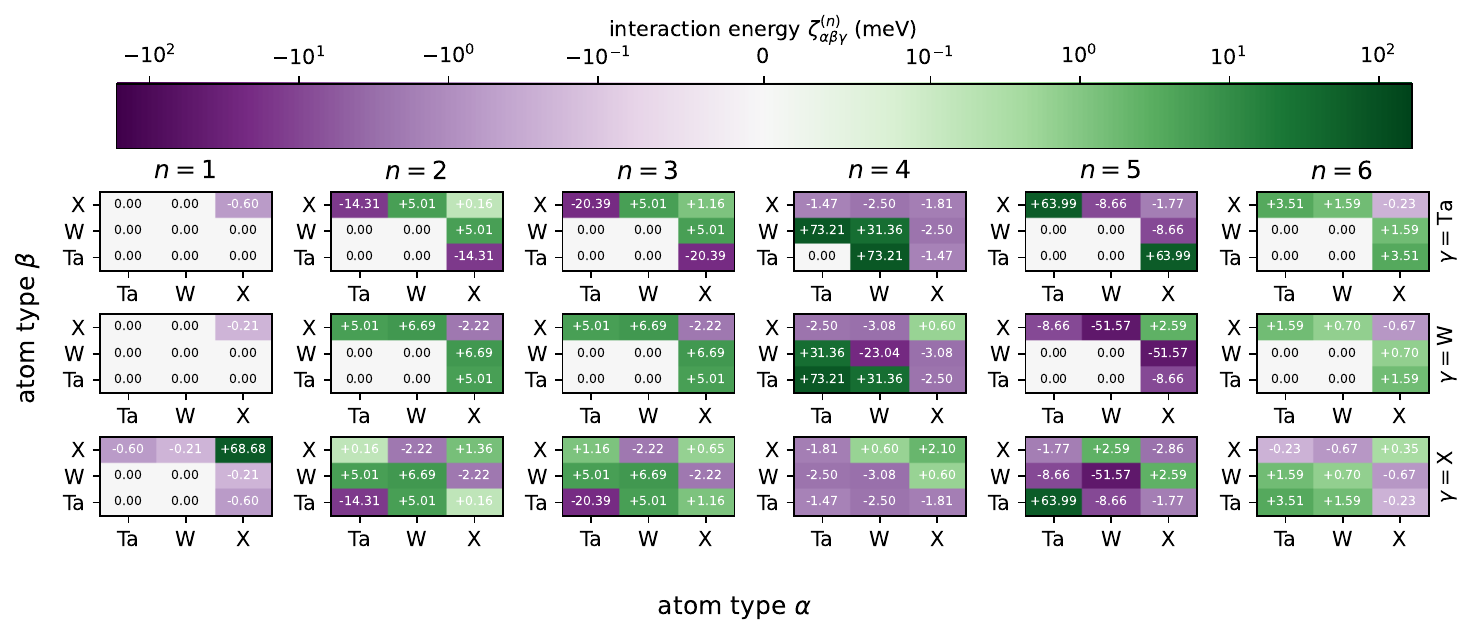}
    \caption{Pairwise interaction coefficients $\boldsymbol{\varepsilon}$ (top) and three-body interaction coefficients $\boldsymbol{\zeta}$ (bottom). For the pairwise coefficients, each heatmap is labeled with an interaction order $n$, with each cell corresponding to a given $\alpha$-$\beta$ pair. For the three-body coefficients, each heatmap is labeled with an interaction order $n$ as well as an atom type $\gamma$, with each cell corresponding to a given $\alpha$-$\beta$ pair. In both cases, the purple shades represent attraction, while the green shades represent repulsion, with shade corresponding to strength of the interaction.}
    \label{fig:ecis}
\end{figure}

We find that our fitted coefficients respect important physical constraints. Firstly, non-physical interactions have a resulting fitted interaction energy of $0$, e.g. $\varepsilon_{\text{W}\text{W}}^{(1)}$, which describes the non-physical interaction between a W atom at a transition state site and a W atom at a nearby ordinary lattice site. Secondly, the learned coefficients respect the symmetries of each cluster type, i.e. two features that are symmetrically equivalent have the same interaction energy. This manifests as symmetry within the tensors, i.e., given any permutations $\sigma\in S_2$ and $\tau\in S_3$, the fitted interaction tensors respect the symmetry $\varepsilon_{\alpha\beta}^{(n)} = \varepsilon_{\sigma(\alpha\beta)}^{(n)}$ and $\zeta_{\alpha\beta\gamma}^{(n)} = \zeta_{\tau(\alpha\beta\gamma)}^{(n)}$, where $S_k$ denotes the symmetric group on $k$ elements. Thirdly, we find that the learned two-body interaction coefficients are larger in magnitude than the learned three-body interaction coefficients. Furthermore, a small set of coefficients dominate the model. Namely, of the full set of unique clusters that satisfy the one-hot constraint, which has size $\text{rank}(\Delta\mathbf{T}) = 56$, only 21 coefficients (that are unique up to symmetry) are larger than $\SI{10}{meV}$ in magnitude (Fig.~\ref{fig:ecis}).

Our fitting routine must additionally respect the constraint that a feature difference $\Delta\mathbf{t} = \mathbf{0}$ should identically evaluate to an energy difference $\Delta E = 0$. We achieve this by ensuring that our routine fits a model that does not include any intercept.

\begin{figure}[H]
    \centering
    \includegraphics[width=\linewidth]{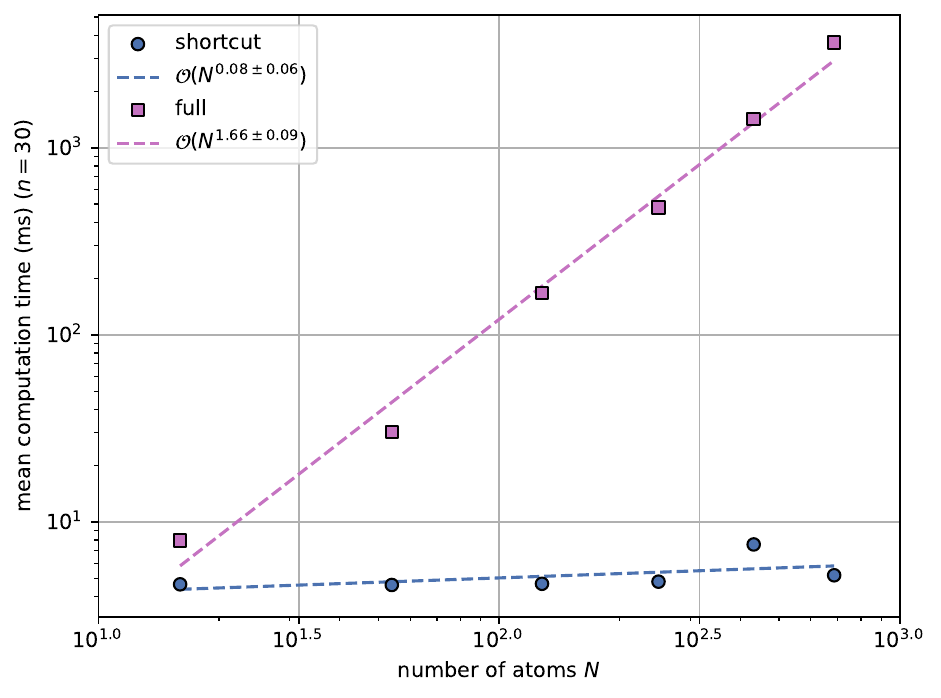}
    \caption{Computation time of energy barriers using Eq.~\ref{eq:feat-vec-diff} (labeled shortcut) compared to independently predicting two energies $E^\ddagger$ and $E^\circ$ from the configurations $\mathbf{X}^\ddagger$ and $\mathbf{X}^\circ$. Each point is computed as the mean of $n = 30$ randomly generated hops in equiatomic Ta-W. Included are linear fits along the log-log scale, and corresponding power laws.}
    \label{fig:benchmark}
\end{figure}

In terms of model performance, we find that the fitted model, similar to our prior work \cite{jeffries2026generalized}, predicts energy barriers significantly more efficiently when using feature differences $\Delta\mathbf{t}$, i.e. Eq.~\ref{eq:feat-vec-diff}, rather than independently predicting two energies $E^\ddagger$ and $E^\circ$ from the configurations $\mathbf{X}^\ddagger$ and $\mathbf{X}^\circ$ as done by Li et al. \cite{li2021predicting}. Specifically, only including sites that are local to the event, i.e. sites in $\mathcal{D}$, offers a roughly quadratic speedup in energy barriers with respect to system size (Fig.~\ref{fig:benchmark}).

\begin{figure}[H]
    \centering
    \includegraphics[width=\linewidth]{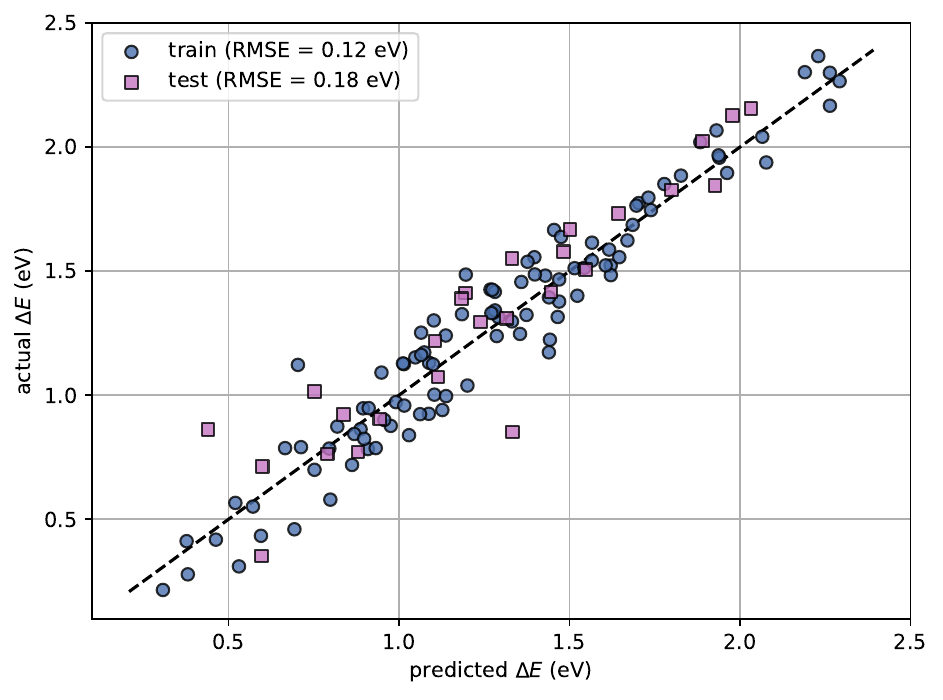}
    \caption{Parity plot of energy barrier predictions.}
    \label{fig:fitting}
\end{figure}

Barring a small set of outliers, we additionally see that our model predicts the energy barriers quite well. Notably, the model performs similarly on the training and testing set, and residuals are fairly uniformly distributed about zero error. So, our model is likely not overfitted, and is likely adequately posed, i.e. it contains a sufficient number of clusters to accurately describe our energy barriers (Fig.~\ref{fig:fitting}).

Lastly, the root mean square error (RMSE) in the training and testing sets, respectively $\SI{0.12}{eV}$ and $\SI{0.18}{eV}$, are relatively small, but non-negligible. We expect this to have minimal effect on our final calculations, namely the activation energy for vacancy diffusion, but can be potentially problematic for other computations. We have elaborated on this limitation in Sec.~\ref{sec:barrier-limitation}. Furthermore, these errors should be easily mitigated simply by increasing the feature space dimension, i.e. the number of considered clusters. However, our multi-lattice introduces a large feature space, namely $234$ features, even for the relatively small maximum neighbor distance used in this study of $\sqrt{5}a/2 \approx 1.12a$, so introducing even more features runs a large risk of overfitting. However, many of these features are highly correlated, or even identical, suggesting that the fitting problem is well-addressed by typical dimensionality-reduction techniques such as principal component analysis (PCA). We have an exploration of such techniques for a future work.

Using our model for the energy barriers, we can simulate vacancy diffusion in varying compositions and temperatures, namely between $5\%$ and $95\%$ Ta and $\SI{600}{K}$ and $\SI{3000}{K}$, and calculate diffusivity as a function of both composition and temperature. To do this, we set up a $5\times 5\times 5$ supercell of our multi-lattice using the Atomic Simulation Environment library \cite{ase-paper}, and initialize one vacancy on the bcc sublattice, with energy $E^\circ$. From this initial configuration, we identify the nearest atomic neighbors of the initialized vacancy. We then compute the possible transition states by moving the nearest atomic neighbors to the transition state sites sitting between them, with energies $E_i^\ddagger$.

Then, using \verb|tce-lib|, we compute the feature difference between the two states, i.e. $\Delta_i\mathbf{t} = \mathbf{t}_i^\ddagger - \mathbf{t}^\circ$. This is computed using the same local difference shortcut as in our prior work \cite{jeffries2026generalized}, which calculates feature differences solely locally, avoiding calculating redundant cluster difference counts far away from the hop. From these feature differences, we then predict the energy barrier $\Delta_i E = \mathbf{j}^\intercal\Delta_i\mathbf{t}$ from our fitted model above. Then, according to HTST, each rate is:

\begin{equation}
    R_i = \nu \exp\left(-\frac{\Delta_i E}{k_BT}\right)
\end{equation}

where $\nu$ is the rate prefactor. In this work, we set this to a constant $\nu = \SI{1e+13}{Hz}$. Then, we choose the hopping event with probability $p_i$ according to the residence-time algorithm \cite{bortz1975new}:

\begin{equation}
    p_i = \frac{R_i}{\sum_j R_j}
\end{equation}

and we evolve the time by sampling the exponential distribution of the effective rate:

\begin{equation}
    \Delta t \sim \text{Exp}\left(\sum_i R_i\right)
\end{equation}

\begin{figure*}
    \centering
    \includegraphics[width=\linewidth, height=0.45\textheight, keepaspectratio]{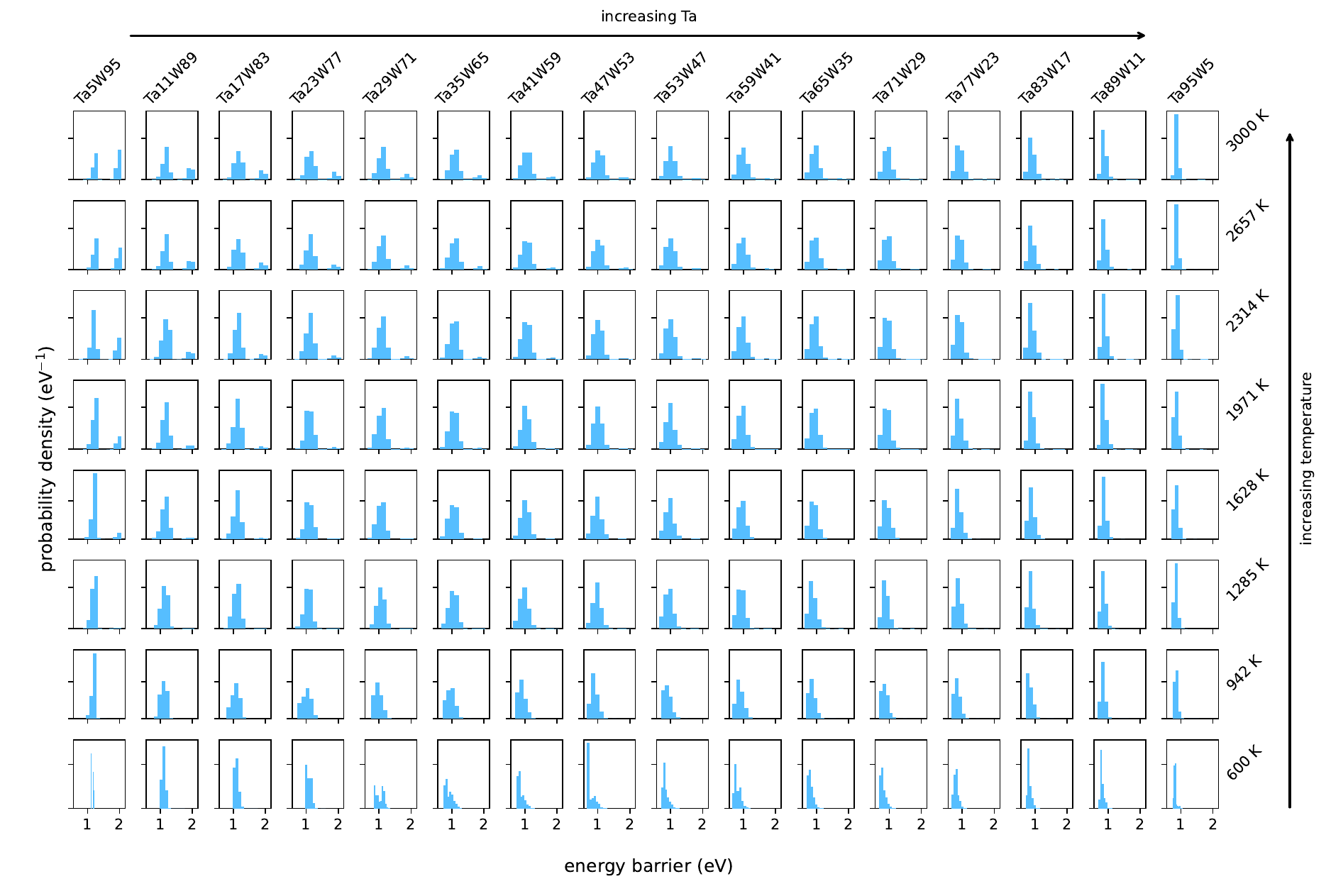}
    \caption{Distributions of traversed barriers within the KMC trajectories for each tested composition and temperature.}
    \label{fig:barrier-distributions}
\end{figure*}

We then repeat this time-evolution step for $500,000$ steps per temperature and composition. From each trajectory, we see interesting dynamics in regards to the distribution of barriers traversed over time. Namely, there seems to be a boundary in composition-temperature space in which the vacancy traverses a bimodal distribution of energy barriers, rather than a unimodal one as one might expect for a system like Ta-W, where the elements present in the solution have a large degree of chemical similarity (Fig.~\ref{fig:barrier-distributions}).

From the distribution of traversed barriers, we see that there are two main peaks emerging as a function of composition and temperature. For Ta-rich environments, the lower-barrier peak becomes more prominent over the high-barrier peak. This is consistent with the observation that the energy barrier for vacancy diffusion is lower in pure Ta than in pure W. From our fitted model, we have computed these barriers to be respectively $\Delta E_\text{Ta} = \SI{0.84}{eV}$ and $\Delta E_\text{W} = \SI{1.97}{eV}$.

Interestingly, though, this relative prominence is not symmetric over composition space. For example, at high temperatures in Ta-rich solutions, there is nearly no W peak. However, even in W-rich solutions, the Ta peak is quite prominent. This asymmetry is even stronger at lower temperatures, in which the W peak is not strongly present over most of the composition range. This suggests that, even at high temperatures, the vacancy is biased towards hopping in relatively Ta-rich environments, even for low Ta concentration. 

\begin{figure}[H]
    \centering
    \includegraphics[width=\linewidth]{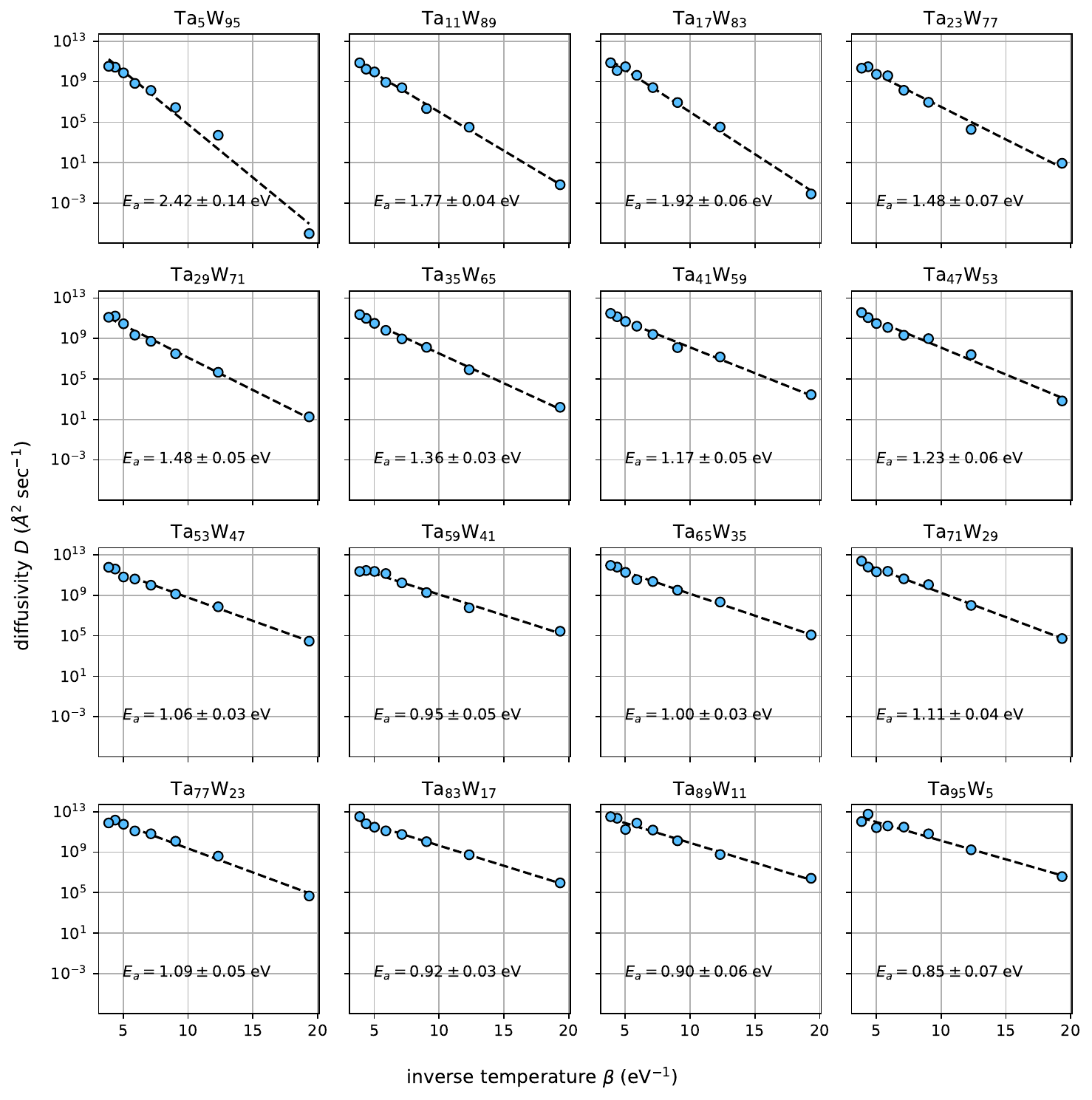}
    \caption{Predicted vacancy diffusivity as a function of composition and temperature in the Ta-W system.}
    \label{fig:diff}
\end{figure}

From the KMC trajectories, we then compute the mean-squared displacement (MSD), for each composition and temperature, of the vacancy as a function of time $t$ with an efficient algorithm developed by Calandri et al. \cite{calandrini2011nmoldyn} that utilizes the fast Fourier transform. Then, from the MSD curves, we utilize the Einstein relation for diffusivity:

\begin{equation}
    D = \lim_{t\to\infty}\frac{\left\langle \left \| \mathbf{r}(t_0 + t) - \mathbf{r}(t_0)\right\|^2\right\rangle_{t_0}}{6t}
\end{equation}

where $\langle\cdot\rangle_{t_0}$ denotes an average over possible time segments starting at $t_0$, and $\mathbf{r}(t)$ is the position of the vacancy at time $t$. In practice, though, there are a diminishing amount of possible trajectory segments of size $t$ as $t$ grows. To address this, we calculate the MSD for the entire trajectory, but linearly fit the MSD as a function of time for the first half of the data. Then, the slope of this fit is $6D$. We see that each diffusivity is well-described by the Arrhenius relation (Fig.~\ref{fig:diff}):

\begin{equation}
    D(x, T) = D_\infty(x)\exp\left(-\frac{E_a(x)}{k_BT}\right)
\end{equation}

where $E_a(x)$ is the composition-dependent activation energy for diffusion, and $D_\infty(x)$ is the composition-dependent prefactor. 

\begin{figure}[H]
    \centering
    \includegraphics[width=\linewidth]{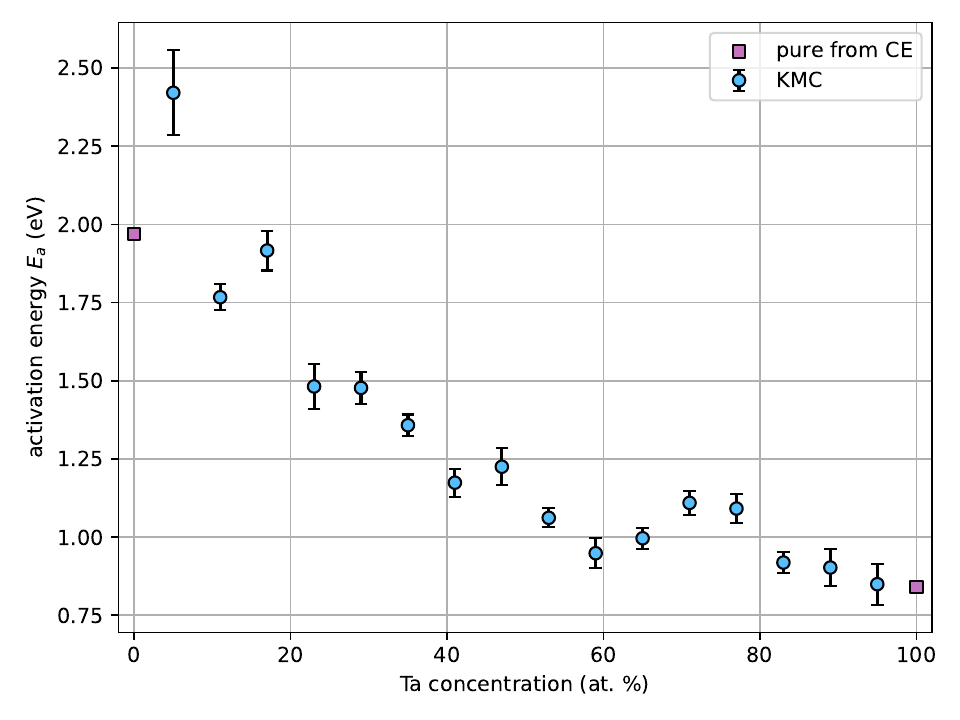}
    \caption{Predicted activation energy for vacancy diffusion as a function of composition in the Ta-W system.}
    \label{fig:energies}
\end{figure}

From this fit, we see that the effective energy barrier roughly decreases linearly with Ta concentration for moderately concentrated solutions. This is unsurprising due to the chemical similarity between Ta and W stated above, making a Vegard's law-like expression plausible (Fig.~\ref{fig:energies}).

However, locally, there are interesting regimes. Perhaps the most interesting is the dilute Ta region, in which dilute Ta content seems to increase the activation energy for diffusion, rather than decrease it. We attribute this to the strong preference for the vacancy to hop in Ta-rich environments, even in the W-rich region of composition space, at high temperatures, as observed in our barrier hopping distributions (Fig.~\ref{fig:barrier-distributions}). This is consistent with experimental work by Ipatova et al. \cite{ipatova2019void}, in which they observed suppressed void growth with the addition of $\SI{5}{wt. \%}$ Ta to W at $\SI{800}{^\circ C}$, indicating hindered vacancy mobility. Our results are also consistent with thermodynamic calculations performed by Pandey et al., who showed, using the CE technique, that vacancies preferentially bind with Ta in the Ta-W system, hypothesizing that Ta clouds should slow down vacancy migration \cite{pandey2025synergistic}. More broadly, this is also consistent with prior descriptions of solute drag in dilute bcc alloys, in which vacancies can orbit their first nearest-neighbor sites long enough to impede long-range diffusion \cite{garnier2013solute, messina2014exact, messina2016systematic}. However, beyond the dilute regime, i.e. for more than $11\%$ Ta, we find that the vacancy's diffusion is no longer impeded by Ta-rich orbits. In essence, when Ta is small enough in concentration, Ta can form small islands in which the vacancy orbits. Then, once Ta is large enough in concentration, these islands can percolate, supporting long-range diffusion that is enhanced by Ta.

This is consistent with the well-known bond percolation threshold in bcc solids in 3D, $p_c\approx 0.18$ \cite{lorenz1998precise}. Namely, in a random solution, the number of Ta atoms that neighbor the vacancy is binomially distributed, i.e. $N_\text{Ta}\sim \mathcal{B}(z, x_\text{Ta})$, where $z = 8$ is the coordination number of the lattice. Then, if we consider an environment to be Ta-rich if there are at least $n$ nearest neighbor Ta atoms, we can find the region of composition space where Ta forms a percolated network by solving the inequality $P(N_\text{Ta}\geq n) \geq p_c$:

\begin{equation}
    \begin{aligned}
        P(N_\text{Ta}\geq n) &= 1 - \sum_{k=0}^{n-1}P(N_\text{Ta} = k) \\
        &=1 - \sum_{k=0}^{n-1}\binom{z}{k}x_\text{Ta}^k (1-x_\text{Ta})^{z-k}\\
        &\geq p_c
    \end{aligned}
\end{equation}

\begin{figure}[H]
    \centering
    \includegraphics[width=\linewidth]{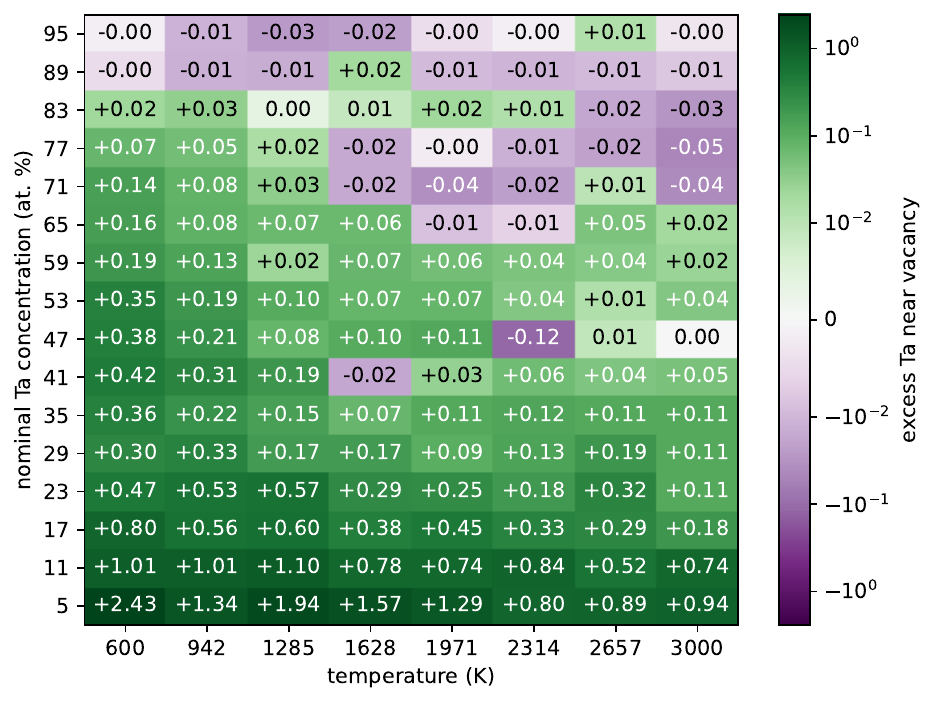}
    \caption{Ta excess near vacancy $\Gamma_\text{Ta}$ throughout the KMC simulations. Each cell represents a simulation at a fixed temperature and nominal composition. The purple shades represent attraction of the vacancy to Ta, i.e. excess Ta, while the green shades represent repulsion, i.e. depleted Ta, with shade corresponding to magnitude of excess/depletion.}
    \label{fig:excess}
\end{figure}

For $n = 2$, this yields a critical concentration $x_\text{Ta}^*\approx 9.8\%$, i.e. if $x_\text{Ta} \geq 9.8\%$, then Ta forms percolated networks. This is very consistent with our predicted effective activation energies, in which there is both a sharp drop in activation energy between $5\%$ and $11\%$ Ta, as well as a relatively large increase in activation energy in $5\%$ Ta relative to pure W. Furthermore, the standard error for the activation energy at $5\%$ Ta is much larger than for other compositions, which is suggestive of a phase transition. From the KMC trajectories, we additionally find that the average excess Ta near the vacancy generally increases with W concentration, with more excess at lower temperatures (Fig.~\ref{fig:excess}). We define this excess as:

\begin{equation}
    \Gamma_\text{Ta} = \frac{\langle \phi_\text{Ta}\rangle - x_\text{Ta}}{x_\text{Ta}}
\end{equation}

where $\langle \phi_\text{Ta}\rangle$ is the average fraction of Ta atoms within the local environment of the hopping vacancy, which we take to be the first and second nearest neighbor shells. Then, if $\Gamma_\text{Ta} > 0$, the vacancy on average spends more time near Ta atoms than if the vacancy had no bias towards orbiting near any atom type.

The results for excess Ta near the vacancy are consistent with the percolation picture, i.e., for small Ta concentrations, the vacancy is biased towards in Ta-rich environments, inhibiting diffusion.

We find that dilute W does not have a similar effect in pure Ta. Namely, dilute W does not significantly change the activation energy for vacancy diffusion. Within this regime, the distributions of visited energy barriers are very similar, with no visible W peak until about 60\% W at high temperatures.

Then, in the regime where Ta is concentrated enough to form percolated networks but the alloy remains majority W ($11\%$ to $59\%$ Ta), we find that Ta enhances diffusion. We attribute this to something more trivial, wherein Ta enhances diffusion from energetics alone, but W is sufficiently present such that hops in W-rich environments are no longer avoidable, yielding a regime in which a mean-field-theory picture is appropriate. This can also be seen in the traversed barrier distributions, in which there is a smaller, but significant W-peak at high temperatures within this composition range.

%% file: sections/discussion.tex
\section{Discussion}

The ability to accurately predict diffusion in concentrated alloys remains a central challenge in computational materials science, owing to the strong dependence of migration barriers on the local chemical environment. Traditional approaches often rely on simplified kinetic models, such as the KRA approximation, or assume a small number of representative barriers, which can obscure the role of local configurational variability in governing transport.

From a practical standpoint, this framework enables predictive first-principles modeling of diffusion in multicomponent alloys without relying on ad hoc kinetic approximations, extending the framework of Li et al. to model more diffusion mechanisms than a simple first-nearest-neighbor vacancy hop. In particular, it opens the door to high-throughput screening of alloy compositions with tailored transport properties with a wide range of diffusive mechanisms, as well as the investigation of complex phenomena such as solute drag, percolation of fast diffusion pathways, and composition-dependent activation energies. More broadly, the methodology is general and can be extended to other defect-mediated processes, including self-interstitial diffusion and defect clustering, provided that the relevant transition-state energetics are adequately sampled. As such, this work establishes a physically consistent and rigorous framework for bridging first-principles calculations and mesoscale kinetics in chemically complex materials, with potential applications spanning alloy design, irradiation response, and high-temperature materials performance.

Beyond predictive capability, though, the present framework also enables new avenues for engineering diffusion in concentrated alloys. By explicitly resolving the distribution of migration barriers and their dependence on local chemical environments, diffusion can be understood in terms of the connectivity of low-barrier pathways on the lattice. This provides a natural route to tuning transport via percolation: alloy compositions may be designed to either promote or suppress the formation of connected networks of fast diffusion channels, thereby controlling overall mobility. 

Similarly, the framework provides a quantitative description of solute drag as an emergent consequence of local barrier statistics, enabling the identification of compositions that inhibit or enhance vacancy motion. These effects have direct implications for high-temperature performance and radiation response, where vacancy mobility governs processes such as creep and void growth. In particular, suppressing the percolation of low-barrier pathways or enhancing trapping environments offers a potential strategy for reducing vacancy flux and mitigating void formation. For applications where we might be interested in inhibiting diffusion, e.g. minimizing void growth, this work then suggests that dilute Ta is optimal. On the other hand, for applications where we might be interested in enhancing diffusion, e.g. increasing vacancy flux towards a defect sink to annihilate vacancies, one can then engineer percolated networks simply by having a Ta concentration larger than $x_\text{Ta}\approx 9.8\%$. This drag effect also indicates that, in the Ta-W system, Ta should segregate near point defect sinks. We additionally hypothesize that this dragging/percolation effect should hold for other solutes in which vacancy migration barriers are locally lowered. For example, first-principles studies show that Re and Os have lower vacancy-mediated diffusion barriers in W than W's self diffusion barrier \cite{suzudo2014stability, li2017behaviors, li2019radiation}. This effect could therefore be relevant in studying how kinetically-controlled properties of plasma-facing W armor in fusion reactors, in which transmutation via neutron irradiation yields roughly $3.8\%$ Re and $1.4\%$ Os in pure W over a period of five years \cite{gilbert2011neutron}, evolve over the lifetime of the materials in irradiated environments. 

More broadly, by directly learning environment-dependent migration barriers from first-principles data, the present approach avoids the need for reduced kinetic models such as the KRA approximation. This enables a unified and systematically improvable framework for modeling diffusion across a wider class of chemically complex materials, where local environments and transition-state energetics cannot be captured by simplified parameterizations, or where modeling a larger class of diffusive mechanisms is crucial, e.g. SIA diffusion and/or rotation under irradiation.

%% file: sections/limitations.tex
\section{Limitations and future work}

Migration rates in this work are evaluated within the framework of HTST, in which the activation free energy is approximated by the difference in potential energy between the transition and initial states. This approximation is formally valid in the low-temperature limit, where vibrational free energy contributions are assumed to be harmonic and largely cancel between configurations. At elevated temperatures, however, entropic effects can become non-negligible. In particular, vibrational entropy differences between the initial and transition states, as well as anharmonic contributions, can lead to significant deviations from Arrhenius behavior. Recent studies in W have demonstrated that migration free energies exhibit strong temperature dependence due to anharmonic vibrational effects, resulting in measurable non-Arrhenius diffusion behavior \cite{zhang2025ab}. These results suggest that a purely energetic description of migration barriers may neglect important finite-temperature contributions.

Nevertheless, the cluster expansion formalism employed here can, in principle, be extended to incorporate vibrational contributions. For example, temperature-dependent effective cluster interactions may be constructed by fitting to free energies rather than energies, enabling the inclusion of vibrational entropy within the same expansion framework \cite{levesque2011simple, martinez2012decomposition}. Such extensions provide a systematic route toward incorporating finite-temperature effects beyond the HTST approximation.

Additionally, the KMC simulations performed in this work sample configurations that are, to a good approximation, representative of a random solid solution. While local compositional fluctuations naturally arise from vacancy-mediated dynamics, the system does not exhibit significant evolution toward ordered states over the timescales considered. As such, the underlying ensemble remains close to a disordered configuration throughout the simulations. In general, SRO can have a nontrivial impact on diffusion kinetics by modifying the local chemical environment encountered by migrating defects. Changes in SRO alter the distribution of migration barriers, potentially leading to variations in effective diffusivities and activation energies. In particular, the emergence of preferred local motifs may either facilitate or hinder vacancy motion depending on the associated energetics.

However, the temperatures considered in this study lie well above the order-disorder transition temperature (ODTT) of the Ta-W system, which has been estimated to fall in the range of $\SI{200}{K}$ to $\SI{400}{K}$ \cite{sun2025thermodynamic}. Therefore, for our temperature range ($\SI{600}{K}$-$\SI{3000}{K}$), equilibrium configurations are expected to exhibit minimal SRO, and the assumption of a random solid solution is therefore well justified. Consequently, the influence of ordering on the computed diffusion properties is expected to be limited under the present conditions.

We note, however, that at lower temperatures approaching the ODTT, or in systems with stronger ordering tendencies, SRO could play a more significant role in determining diffusion behavior. In such cases, a fully coupled treatment of thermodynamics and kinetics would be required to accurately capture the interplay between ordering and transport. This is especially relevant for studying percolation effects, in which it has been recently shown that percolation thresholds can strongly depend on chemical SRO in binary alloys \cite{yu1994correlated, frary2007correlation, roy2024effect, ROY2026117137}.

Thirdly, the KMC simulations in this work are performed on finite supercells (bcc $5\times5\times5$), which may introduce finite-size effects. In particular, small systems may suppress long-wavelength compositional fluctuations or introduce weak correlations through periodic boundary conditions. Additionally, the diversity of local environments sampled by the vacancy may be reduced in smaller supercells, potentially affecting the statistics of rare configurations. That said, such effects are typically most pronounced in systems involving long-range interactions, defect clustering, or spatially extended phenomena.

For the present case of dilute vacancy diffusion in a disordered solid solution, where kinetics are governed primarily by local environments and short-range energetics, these finite-size effects are expected to be limited. Nevertheless, systematic finite-size scaling could be employed in future work to further quantify any residual size dependence of the computed diffusivities.

Lastly, the KMC simulations in this work are performed in the dilute-defect limit, with a single vacancy evolving within the simulation cell. As such, interactions between multiple vacancies, including vacancy-vacancy binding and clustering, are not explicitly considered. These effects can become important at elevated defect concentrations, where the formation of vacancy clusters, voids, or extended defects may kinetically trap vacancies and, in turn, affect mechanical properties such as swelling, creep, and irradiation response. In the present context, the single-vacancy approximation enables a focused investigation of the local chemical factors governing migration barriers and diffusivity in a disordered solid solution. For sufficiently low equilibrium vacancy concentrations, this approximation is expected to provide an accurate description of tracer-like diffusion behavior.

We note, however, that the framework developed here is not inherently restricted to the single-defect limit. In principle, the same training set could be augmented to include multiple vacancies, providing vacancy-vacancy and higher-order defect interactions within the training data. Such an approach would enable the study of vacancy clustering kinetics and collective diffusion phenomena within a consistent atomistic framework.

%% file: sections/conclusions.tex
\section{Conclusions}

In this work, we have extended our prior work on TCE models by augmenting the lattice with transition state sites, embedding transition states for vacancy migration into configuration space, allowing us to treat transition states of a large class of diffusive mechanisms as distinct configurations within configuration space. We then use this construction to fit vacancy migration barriers in the Ta-W alloy, showing that this allows us to accurately predict the energy barrier of a vacancy hop in Ta-W as a function of features local to the hop without the use of the KRA relation.

We then use this fitted model to simulate vacancy diffusion using a KMC methodology. From the resulting diffusion data, we argue that vacancy diffusion in this system follows 3 distinct regimes:

\begin{itemize}
    \item Dilute Ta ($0\% \leq x_\text{Ta} \leq 5\%$), in which Ta acts as a dragging solute
    \item Moderate Ta ($11\% \leq x_\text{Ta} \leq 59\%$), in which Ta still dominates diffusion, but forms percolated networks that support long-range diffusion
    \item Moderate to dilute W ($64\% \leq x_\text{Ta} \leq 100\%$), in which the vacancy avoids W, exhibiting diffusion similar to pure Ta
\end{itemize}

More generally, this work enables the combination of first-principles transition-state energetics with data-driven cluster expansions beyond the KRA approximation in a way that is generalizeable to a wide class of diffusive mechanisms, further enabling a physically transparent and computationally efficient framework for predicting diffusion in complex alloys, providing a foundation for both mechanistic understanding of diffusion, and therefore materials design, for advanced applications.

%% file: sections/data_availability.tex
\section{Data Availability}

Excluding the raw VASP and NEB data, all data and code necessary to replicate this study is available on GitHub \cite{muexly_taw_tce_kmc}. Raw VASP data has been excluded to avoid copyright issues, but will be made available upon reasonable and legal request.

%% file: sections/acknowledgements.tex
\section{Acknowledgements}

Authors acknowledge support from the U.S. Department of Energy, Office of Basic Energy Sciences, Materials Science and Engineering Division under Award No. DE-SC0022980.

Additionally, this material is based on work supported by the National Science Foundation under Grant Nos. MRI\# 2024205, MRI\# 1725573, and CRI\# 2010270 for allotment of compute time on the Clemson University Palmetto Cluster.

%% file: sections/disclaimer.tex
\section{Disclaimer}

Any opinions, findings, and conclusions or recommendations expressed in this material are those of the author(s) and do not necessarily reflect the views of the National Science Foundation.

%% file: sections/appendix.tex
\appendix

\section{Appendix}

\subsection{Energy barrier error propagation}\label{sec:barrier-limitation}

Suppose the true energy barrier is $\Delta E^\circ$, but we predict an energy barrier $\Delta E = \Delta E^\circ + \varepsilon$, where $\varepsilon\sim\mathcal{N}(0, \sigma^2)$ is normally distributed. We then predict a rate, on average:

\begin{equation}
    \begin{aligned}
        \langle\hat{r}\rangle &= \nu \exp\left(-\frac{\Delta E^\circ}{k_BT}\right)\left\langle\exp\left(-\frac{\varepsilon}{k_BT}\right)\right\rangle\\
        &=r_\text{true}\exp\left(\frac{\sigma^2}{2(k_BT)^2}\right)
    \end{aligned}
\end{equation}

where $r_\text{true} = \nu \exp\left(-\frac{\Delta E^\circ}{k_BT}\right)$ is the true rate and $\langle \hat{r}\rangle$ is the average predicted rate. Then, the error blows up exponentially, both with inverse temperature and prediction variance $\sigma^2$. For test RMSE $\SI{0.2}{eV}$ at temperature $\SI{500}{K}$, this amounts to roughly $\exp\left(\frac{\sigma^2}{2(k_BT)^2}\right) \approx 5\times 10^4$, i.e. $4$-$5$ orders of magnitude difference. We expect that this has minimal effect on computing quantities like the activation energy, which are proportional to $\ln\hat{r} \propto -\frac{\varepsilon}{k_BT}$. However, for computing absolute diffusivities, it might be necessary to estimate rates with a different strategy. Two possible solutions might be correcting the predicted rates by $\exp\left(\frac{\sigma^2}{2(k_BT)^2}\right)$, or by fitting rates directly rather than energy barriers. We save an exploration of these strategies for a future work.